\documentclass[12pt]{article}
\usepackage{hyperref}
\usepackage{times,amssymb,euscript,epsfig,latexsym,amsmath}
\usepackage{graphicx}
\begin{document}
\newcommand{\todo}[1]{\vspace{5 mm}\par \noindent
\framebox{\begin{minipage}[c]{0.95 \textwidth}
\tt #1 \end{minipage}}\vspace{5 mm}\par}
\newtheorem{thm}{Theorem}[section]
\newtheorem{prop}[thm]{Proposition}
\newtheorem{lem}[thm]{Lemma}
\newtheorem{cor}[thm]{Corollary}
\newtheorem{dfn}[thm]{Definition}
\newtheorem{proof}{Proof}[section]
\date{}
\title{ \Large Poisson brackets for the dynamically coupled system of a free boundary and a neutrally buoyant rigid body in a body-fixed frame.}
\author{ Banavara N. Shashikanth\footnote{Mechanical and Aerospace Engineering
Department, MSC 3450, PO Box 30001, New Mexico State University,
Las Cruces, NM 88003, USA. E-mail:shashi@nmsu.edu}}
\maketitle

\begin{abstract}
The fully coupled dynamic interaction problem of the free surface of an incompressible fluid and a rigid body beneath it, in an inviscid, irrotational framework and in the absence of surface tension, is considered. Evolution equations of the global momenta of the body+fluid system are derived. It is then shown that, under fairly general assumptions, these evolution equations combined with the evolution equation of the free-surface, referred to a body-fixed frame, is a Hamiltonian system. The Poisson brackets of the system are the sum of the canonical Zakharov bracket and the non-canonical Lie-Poisson bracket. Variations are performed consistent with the mixed Dirichlet-Neumann problem governing the system. 
\end{abstract}

\newpage 
\tableofcontents

\newpage

\section{Introduction.}

  Water wave dynamics may be described as both a classical and modern topic of research in fluid dynamics. Lamb's book \cite{La1932} (Chapter VIII on `Tidal Waves') contains references to several important papers of the classical oeuvre. The relation of the Korteweg-de Vries (KdV) equation to integrability and soliton theories has been a relatively more recent development \cite{GaGrKrMu1967} but has spawned a very large and  active field of research \cite{FaTa07}. Johnson's book \cite{Jo1997} provides a nice introduction to both classical and modern aspects of the subject of water waves. 

      Theoretical investigations of rigid bodies in water also has a rich history, in particular, the topic of bodies floating on the water surface; some landmark publications being \cite{Jo1949, Jo21949, Ur1948, Ur21953, We1971}. Most approaches to these problems--with applications to ship and marine vehicle motions--- are typically in a linearized framework or/and with the body executing prescribed motions, or with the objective of deriving expressions for the hydrodynamic loads on the body \cite{SaTuFa1970}. The fully coupled nonlinear problem has not been as well investigated theoretically. 
Most investigations of the nonlinear problem are numerical; see, for example, \cite{WuTa2003} and references therein. 

  Papers on Lagrangian and Hamiltonian formulations of the coupled problem are even fewer. Miloh presented Lagrangian formulations in cases where the body is executing oscillations or is set in impulsive motion, on or below the free surface \cite{Mi1984, Mi2001}. Miloh also derived expressions for the hydrodynamical reaction forces on the body. To the best of the author's knowledge, the Hamiltonian formulation of the fully coupled system was first presented by van Daalen, van Groesen and Zandbergen \cite{Da1993, DaGrZa1993}. Working in a spatially fixed frame and without any explicit reference to Poisson brackets, they showed that the combined system is a canonical Hamiltonian system, with the Hamiltonian being the sum of the fluid and body kinetic+potential energies. 

        In this paper, the Poisson brackets of the combined system in a body-fixed frame are presented for the case when the body is  completely beneath the free surface and surface tension is ignored. It is shown that the brackets are the sum of the Zakharov bracket \cite{Zakharov1968}, written in a body-fixed frame, and the non-canonical Lie-Poisson bracket \cite{MaRa99}. The paper is organized as follows. In Section 2, the setup of the physical problem and some assumptions made are described. The fluid domain has a flat bottom stationary boundary that extends to infinity in all horizontal directions.  In Section 3, the evolution equations for the combined momenta of the system are presented. First these are derived in a spatially-fixed frame, following a traditional momentum balance analysis, without any assumptions on the buoyancy of the rigid body.  As one would expect, the spatial momenta are not conserved. Conservation is obtained by moving the bottom boundary to infinity and assuming neutral buoyancy. The momentum equations are then transformed to a body-fixed frame. The details of the momentum balance analysis are relegated to Appendices A and B. Section 4 is the main section of the paper in which the variables in the body fixed frame are presented, and it is shown how the variations can be performed consistent with the mixed Neumann-Dirchlet boundary-value problem. The equations of the system are derived and shown to be Hamiltonian relative to the brackets above. Section 5 has some future directions for research. 

   Apart from standard assumptions such as the far-field decay rates of the velocity potential function, and existence and uniqueness of solutions of the mixed Dirichlet-Neumann problem, certain other assumptions are made in the paper. The four main ones are those given by equations~(\ref{eq:mass}),~(\ref{eq:angmass}) and~(\ref{eq:eta2decay}), and the invertibility of the mass matrix given by~(\ref{eq:matrixM}).

 \section{Setup.}

  A schematic sketch of the system being considered is shown in Figures 1 and 2. Introduce the notations $\Sigma_f$ for the free boundary of the incompressible fluid, $\Sigma_{b} \equiv \partial B$ for the fluid-body boundary and $\mathcal{S}$ for the stationary {\it flat} bottom (spanning the horizontal directions). Denote the half-space bounded by $\mathcal{S}$ by $\mathbb{R}^{3+}$. The non-compact fluid domain $D\subset \mathbb{R}^{3+}$ therefore has a boundary which is the disjoint union of three pieces $\partial D= \Sigma_b \cup \Sigma_f \cup \mathcal{S}$. The uniform density fields of the rigid body and the fluid are denoted by $\rho_b$ and $\rho_f$, respectively.
\paragraph{Convention for unit normals.} Before proceeding, the convention for the unit normal field on the different boundary components is established. $n_f$ points {\it away} from $D$, $n_b$ points {\it into} $D$ (and away from the body $B$) and $n_s$ also points {\it into} $D$.  

\begin{figure}[h]
\begin{center}
\includegraphics[scale=0.4]{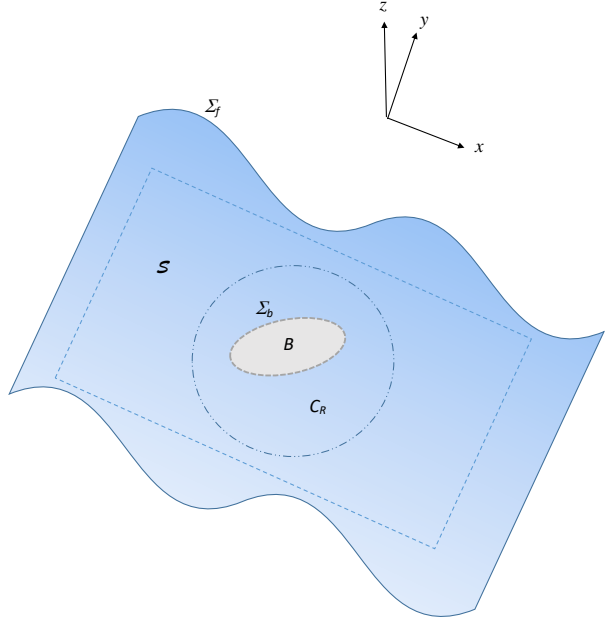}
\end{center}
\vspace{-0.8in}
\caption{Schematic perspective of a rigid body $B$ beneath the free surface $\Sigma_f$ of water. The bottom flat surface $\mathcal{S}$ is shown by the dashed rectangle. Both $\Sigma_f$ and $\mathcal{S}$ extend to infinity in the $x$ and $y$ (horizontal) directions. In the text, the origin of the spatial frame $xyz$ is located at the center of the disc $C_R \subset \mathcal{S}$. }
\label{general}
\end{figure}

\begin{figure}[h]
\begin{center}
\includegraphics[scale=0.5]{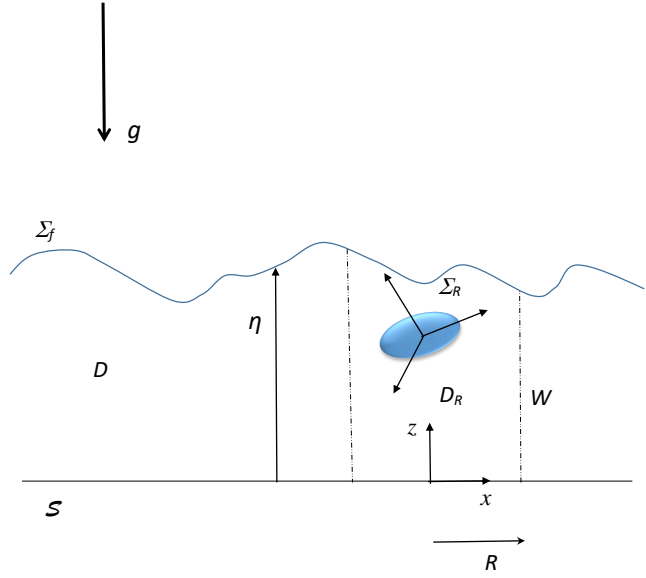}
\end{center}
\vspace{-1.2in}
\caption{A vertical slice of the setup in Figure 1. The body-fixed frame is shown. }
\label{general}
\end{figure}

The function $\Phi$ satisfies the following {\it mixed Dirichlet-Neumann} problem at each time instant $t$: 
\begin{align}
\nabla^2 \Phi &=0 \: {\rm in} \: D , \: \Phi \mid_{\Sigma_f} {\rm prescribed},  \: \nabla \Phi \cdot n_b \mid_{\Sigma_b}= U \cdot n_b, \: \nabla \Phi \cdot n_s \mid_{\mathcal{S}}=0 \nonumber \\
& \hspace{1in}\Phi \rightarrow 0 \: {\rm as \:}  x,y \rightarrow \pm \infty \label{eq:phi} 
\end{align}
where $U$ is the rigid body velocity field in a spatially-fixed frame $xyz$. Both $ \Phi \mid_{\Sigma_f}$ and $U$ are prescribed at initial time but fixed at all later times by the evolution equations. 

\paragraph{Far-field assumptions.} Far from the body, it will be assumed that the fluid surface
is undisturbed and has a constant elevation $\eta_0$. The velocity field also goes
to zero. Since the fluid flow field is vorticity-free, the decay rate of $\Phi$ is assumed to be \cite{Batchelor1967}
\begin{align}
\Phi \sim 1 / \mid r \mid^2 \label{eq:phifar}
\end{align}  Applying
B.E.,
\begin{align*}
\frac{\partial \Phi}{\partial t} + \frac{p}{\rho_f} + \frac{\nabla \Phi \cdot \nabla \Phi}{2} + gz&=f(t)
\end{align*}
to a point on the free surface at far infinity, obtain
\begin{align*}
f(t)&= \frac{p_{atm}}{\rho_f} + g \eta_0= {\rm constant}
\end{align*}
Using the above, B.E. then gives that the pressure at any point as
\begin{align*}
p&= p_{atm} + \rho_f g \left(\eta_0 -z \right) - \rho_f \left(\frac{\partial \Phi}{\partial t}+ \frac{\nabla \Phi \cdot \nabla \Phi}{2} \right)
\end{align*}
Using~(\ref{eq:phifar}) and cylindrical coordinates $(R,\theta,z)$, one can write the far-field pressure distribution as 
\begin{align}
p(R,\theta,z,t)&=p_{atm} +\rho_f  g \left(\eta_0 -z \right) + A(R,\theta,z,t), \quad  A(R,\theta,z,t)=O(1/R^2) \label{eq:pfarzt}
\end{align}

At the flat bottom $\mathcal{S}(z = 0)$, one obtains
\begin{align}
p&=p_{atm}+\rho_f g \eta_0 -\rho_f \left(\frac{\partial \Phi}{\partial t}+ \frac{\nabla \Phi \cdot \nabla \Phi}{2} \right)_{\mid_{\mathcal{S}}} \label{eq:ps}
\end{align}
and at the free surface $\Sigma_f(z=\eta)$, one obtains
\begin{align}
p&=p_{atm}+\rho_f g \left(\eta_0-\eta \right) -\rho_f \left(\frac{\partial \Phi}{\partial t}+ \frac{\nabla \Phi \cdot \nabla \Phi}{2} \right)_{\mid_{\Sigma_f}} \label{eq:pf}
\end{align} 
Note that surface tension effects are absent for the undisturbed surface. 

    Moreover, the waves at the free surface are assumed to satisfy the following two conditions: 
\begin{align}
\rho_f g \int_{\mathcal{S}} \left(\eta_0 - \eta \right) \; \nu_s &=0 \label{eq:mass}  \\
\rho_f g \int_{\mathcal{S}} r \times  \left(\eta_0 - \eta \right) \; \nu_s &=0, \label{eq:angmass}
\end{align}
where $r$ is a position vector (defined later). The bottom flat surface $\mathcal{S}$ has volume (area) form $\nu_s$ and is taken as the datum for the potential energies, and $\eta$ is the free-surface elevation with respect this datum. The first equation may be obviously interpreted as a conservation of mass condition satisfied by the waves on the free surface. The second equation may be viewed as the angular analog of
~(\ref{eq:mass}). It could perhaps be interpreted as a zero global moment, about the gravity axis, due to the waves. 

\paragraph{Total Energy.} Ignoring surface tension and surface energy, the total energy is the kinetic plus potential energy of the fluid+body system, 
\begin{align}
K.E. + P.E.&= \frac{\rho_f}{2} \left(\int_D \left< \left< \nabla \Phi, \nabla \Phi \right> \right> \; \mu + g  \left(\int_{\mathcal{S}}  \eta^2 \; \nu_s - \int_{\mathcal{S}}  \eta_0^2 \; \nu_s \right)\right) \nonumber \\
& \hspace{2in}  +  \frac{\rho_b}{2} \left< \left<U,U \right> \right>_{\mathbb{R}^6} + (\rho_b-\rho_f) \mathcal{V}_B g  z_c ,  \nonumber \\
&= \frac{\rho_f}{2} \left(\int_{\Sigma_f} \Phi \nabla \Phi \cdot n_f \; \nu - \int_{\Sigma_b} \Phi \nabla \Phi \cdot n_b \; \nu\right. \nonumber  \\ 
& \hspace{1.5in} \left.  
+ g  \left(\int_{\mathcal{S}}  \eta^2 \; \nu_s - \int_{\mathcal{S}}  \eta_0^2 \; \nu_s \right)\right) \nonumber \\
& \hspace{1in}   + \frac{1}{2}   \left< \left<(V,\Omega), \rho_b M_b \cdot  ( V,\Omega) \right>\right> + (\rho_b-\rho_f) \mathcal{V}_B g  z_c, \label{eq:totenergy}
\end{align}
where $\mathcal{V}_B$ is the volume of the rigid body, $U \equiv (V,\Omega)$ is the vector of rigid body velocities, $M_b$ is the body mass tensor and $z_c$ is the elevation of the body centroid with respect to the datum $\mathcal{S}$. 
\paragraph{Remark.} Note that the potential energy of the fluid is relative to the undisturbed state. This is done to subtract the infinite potential energy of the fluid in the undisturbed state which is due to the unbounded domain in the horizontal directions, and {\it irrespective} of the location of the datum.  However, subtracting the rest potential energy still does not guarantee that the fluid potential energy is finite. An additional assumption is needed about the rate at which $\eta$ decays in the horizontal directions:
\begin{align}
\int_{\mathcal{S}}  \eta^2 \; \nu_s - \int_{\mathcal{S}}  \eta_0^2 \; \nu_s  < \infty \label{eq:eta2decay}
\end{align}

\section{Global momentum considerations.}

\paragraph{Fluid Momentum.}  Denote by $P_T \equiv (L_T, A_T )$ the momenta
of the total system, i.e. body+fluid system, and by $P_{Tf} \equiv (L_{Tf},A_{Tf} )$ the contribution to these from
the fluid. Using well-known vector identities (see, for example, \cite{Saffman1992} ) allow the latter to be written as follows. Considering the linear momentum first, 
\begin{align*}
L_{Tf }&:=\lim_{R \rightarrow \infty} \rho_f \int_{D_R} \nabla \Phi \; \mu, \\
&= \rho_f  \int_{\partial D_R}  \Phi \; n \; \nu, \\
& {\rm [n \; outward]}, \\
&=\frac{\rho_f}{2} \int_{\Sigma_b} r \times \left(n_b \times \nabla \Phi \right) \; \nu \\
& \hspace{1in} +\lim_{R \rightarrow \infty} \rho_f  \left(  \int_{\Sigma_R} \Phi n_f \; \nu -  \int_{C_R} \Phi  n_s \; \nu + \int_{W}  \Phi e_R \; \nu \right),
\end{align*}
where, as in Figures 1 and 2,  $D_R \subset D$ is a a vertical cylindrical domain of (varying) height $\eta$ and radius
$R$, and bounding surfaces $\Sigma_b$, $\Sigma_R \subset \Sigma_f$, $C_R \subset \mathcal{S}$ and lateral surface $W$ (with outward normal in the radial direction $e_R$). $C_R$ is a circular disc of radius $R$ and $r$ is position vector measured from the origin of the spatially-fixed frame which is taken, wlog, to lie at the center of $C_R$. To avoid notational clutter, the same symbol $\nu$ is used to denote the volume form on any bounding surface. 

 For the first integral on the right, use is made of the vector identity (see SSKM or Saffman) to write it in a different way from the other terms,
\begin{align}
\frac{1}{2} \int_A r \times \left( n \times \nabla \Phi \right)\; \nu&=- \int_A \Phi n \; \nu, \label{eq:vecidL}
\end{align}

\paragraph{Remark.} Note that as $R \rightarrow \infty$, $W$ recedes {\it uniformly} from the body whereas $\Sigma_R$ and $C_R$ do not.

  From the Far-Field Assumptions $\Phi \rightarrow $ constant as $R \rightarrow \infty$, so that the last integral on the right vanishes in this limit, and the expression reduces to 
\begin{align}
L_{Tf }
&=\frac{\rho_f}{2} \int_{\Sigma_b} r \times \left(n_b \times \nabla \Phi \right) \; \nu+\lim_{R \rightarrow \infty} \rho_f \left(  \int_{\Sigma_R} \Phi n_f \; \nu -  \int_{C_R} \Phi  n_s \; \nu \right), \label{eq:fluidlinmom}
\end{align}

    Similarly, considering angular momentum of the flow about the origin of the spatially-fixed frame, 
\begin{align*}
A_{Tf }&:=\lim_{R \rightarrow \infty}\rho_f \int_{D_R} r \times \nabla \Phi \; \mu, \\
&= \lim_{R \rightarrow \infty} \frac{\rho_f}{2} \int_{\partial D_R} r^2 \left(n \times \nabla \Phi \right) \; \nu, \\
& {\rm [n \; outward]}, \\
&=- \frac{\rho_f}{2} \int_{\Sigma_b} r^2 \left(n_b \times \nabla \Phi \right) \; \nu + \lim_{R \rightarrow \infty} \rho_f \left( \int_{\Sigma_R} r  \times  \Phi n_f \; \nu  - \int_{C_R} r \times \Phi n_s   \; \nu \right.  \\
&\left. \hspace{3in} + \int_{W} r  \times  \Phi e_R  \; \nu \right),
\end{align*}
where for the last three integrals use has been made of the vector identity
\begin{align}
\frac{1}{2} \int_{\partial A} r^2 \left( n \times \nabla \Phi \right)\; \nu&= \int_{\partial A} r \times \Phi n \; \nu,   \label{eq:vecidA}
\end{align}
where $\partial A$ denotes the smooth boundary of a domain $A \subset \mathbb{R}^3$. 
As per the Assumption above, $r=R e_R$ on $W$, so that the integral over $W$ vanishes, leaving  
\begin{align}
A_{Tf }
&=- \frac{\rho_f}{2} \int_{\Sigma_b} r^2 \left(n_b \times \nabla \Phi \right) \; \nu + \lim_{R \rightarrow \infty} \rho_f \left( \int_{\Sigma_R} r  \times  \Phi n_f \; \nu  - \int_{C_R} r \times \Phi n_s   \; \nu  \right), \label{eq:fluidangmom}
\end{align}

\paragraph{Total Momentum.} The total body+fluid
momenta in a spatially-fixed frame is 
\begin{align}
P_T \equiv (L_T,A_T)&=\left(L_{Tf}+L_b, A_{Tf}+A_b \right), \label{eq:totmom}
\end{align} 
where 
\begin{align}
L_b&=M_b V, \quad A_b=r_c \times M_b V + I  \Omega \label{eq:bodymomsp} 
\end{align} are the rigid body momenta, with $r_c$ being the position vector of the center of mass in a spatially-fixed frame and $I$ its moment of inertia tensor in a principal-axes frame.   

 Carrying out a traditional momentum analysis, details of which are presented in the Appendix A, one then obtains the following evolution equations in a spatially-fixed frame, 
\begin{align}
\frac{d \mathcal{L}}{dt}&=- \rho_f \int_{\mathcal{S}} \frac{\nabla \Phi \cdot \nabla \Phi}{2} n_s \; \nu + \left(\rho_f  -\rho_b \right) g \mathcal{V}_Bk, \label{eq:dLdt}\\
\frac{d \mathcal{A}}{dt}&=\int_{\mathcal{S}} r \times \left(- \rho_f  \frac{\nabla \Phi \cdot \nabla \Phi}{2}  \right)\; \nu  +r_c \times \left( \rho_f - \rho_b \right) g \mathcal{V}_Bk, \label{eq:dAdt}
\end{align} 
where 
\begin{align*}
\mathcal{L}&=\frac{\rho_f}{2}  \int_{\Sigma_b} r \times \left(n_b \times \nabla \Phi \right) \; \nu+L_b + \rho_f  \int_{\Sigma_f} \Phi n_f  \; \nu, \\
\mathcal{A}&= - \frac{\rho_f}{2} \int_{\Sigma_b} r^2 \left(n_b \times \nabla \Phi \right) \; \nu + A_b + \rho_f \int_{\Sigma_f} r  \times  \Phi n_f \; \nu
\end{align*}
The contributions to the momentum change come from the presence of the fixed surface $\mathcal{S}$ and the lack of neutral buoyancy of the rigid body. The contribution of $\mathcal{S}$ to the momentum change is represented by the  integrals in~(\ref{eq:dLdt}) and~(\ref{eq:dAdt}). From the Far-field assumptions, it is easily seen that these integrals go to zero as $\mathcal{S} \rightarrow z=-\infty$. To obtain global momentum conservation, one therefore needs to make the following assumptions: 
\paragraph{Assumptions.} (a) The surface $\mathcal{S}$ is at  $z=-\infty$ and (b) the rigid body is neutrally buoyant. 
\paragraph{Special case.} Under the above assumptions, 
\begin{align}
\frac{d \mathcal{L}}{dt}&=0 \label{eq:linmomconsv} \\
\frac{d \mathcal{A}}{dt}&=0 \label{eq:angmomconsv}
\end{align}
Henceforth, the paper will only deal with this special case.

\paragraph{Body-fixed frame.} 
Equations~(\ref{eq:linmomconsv}) and~(\ref{eq:angmomconsv}) are now transformed to a body-fixed frame, with origin at the center of mass of the body and axes aligned with the principal axes, using 
\begin{align}
r&=R(t)\cdot l + r_c(t) \label{eq:rlrel}
\end{align} and the general rule for transforming any vector $a \in \mathbb{R}^3$ located at $r$ in the spatially-fixed frame
\begin{align}
a(r)&=R(t) \bar{a}(l), \label{eq:vectrans}
\end{align}
where $\bar{a}$ is the vector located at $l$ in the body-fixed frame.  Using this~(\ref{eq:rlrel}) can also be written as
\begin{align*}
r&=R(t) \left(l + \bar{r}_c \right)
\end{align*}
Real-valued functions transform as 
\begin{align}
\Phi(r,t)&=\Phi'(l,t), \label{eq:ftrans}
\end{align} etc. It follows that 
\begin{align}
\nabla \Phi(r,t)&=R(t) \nabla_b \Phi'(l,t), \label{eq:gradftrans}
\end{align}
etc . Note that under this orthogonal transformation volume forms are preserved. 

  The details of the transformation are presented in Appendix B. The equations take the form 
\begin{align}
\frac{d \mathfrak{L}}{dt}+ \bar{\Omega} \times \mathfrak{L}&=0, \label{eq:lmomeq}\\
\frac{d \mathfrak{A}}{dt} + \bar{\Omega} \times \mathfrak{A} +  \bar{V} \times \mathfrak{L}&=0 \label{eq:angmomeq}
\end{align}
where 
\begin{align}
\mathfrak{L}&=\frac{\rho_f}{2} \int_{\Sigma_b} l \times \left(\bar{n}_b \times \nabla_b \Phi' \right) \; \nu + \bar{L}_b + \rho_f  \int_{\Sigma_f} \Phi' \bar{n}_f  \; \bar{\nu}, \label{eq:bodyfixedL} \\
\mathfrak{A}&=- \frac{\rho_f}{2} \int_{\Sigma_b} l^2 \left(\bar{n}_b \times \nabla_b \Phi' \right) \; \nu + \bar{I \Omega} + \rho_f  \int_{\Sigma_f} l  \times  \Phi' \bar{n}_f \; \bar{\nu}, \label{eq:bodyfixedA}
\end{align}

\section{Variations and Hamiltonian structure in the body-fixed frame.} 

  The total energy function~(\ref{eq:totenergy}) is now written in terms of the variables in the body-fixed frame, keeping in mind the special case(~(\ref{eq:linmomconsv}) and~(\ref{eq:angmomconsv})) and the associated assumptions. 

  Consider the kinetic energy terms first. 
For $h \equiv (R,r_c)$, let $\Psi_h: \mathbb{R}^3 \rightarrow \mathbb{R}^3$ be the map defined by~(\ref{eq:rlrel}).  
\[\psi_h(l)=r. \] Using relations~(\ref{eq:vectrans}),~(\ref{eq:ftrans}) and~(\ref{eq:gradftrans}), 
\begin{align*}
\int_{\Sigma_f} \Phi(r) \nabla \Phi(r) \cdot n_f (r) \; \nu &=\int_{\Psi_h \left(\bar{\Sigma}_f\right)} \Phi (r) \nabla \Phi (r) \cdot n_f(r) \; \nu , \\
&=\int_{\bar{\Sigma}_b} \Phi(\psi_h(l)) \nabla \Phi (\psi_h(l)) \cdot n_b(\psi_h(l)) \; \bar{\nu}, \\
&{\rm [change \; of \; variables \; Theorem]}, \\
&=\int_{\bar{\Sigma}_b} \Phi' (l) R(t) \nabla_b \Phi' (l) \cdot R(t) \bar{n}_b (l) \; \bar{\nu},  \\
&{\rm [using~(\ref{eq:vectrans}) \; and~(\ref{eq:gradftrans})]}, \\
&=\int_{\bar{\Sigma}_b} \Phi' (l) \nabla_b \Phi' (l) \cdot  \bar{n}_b (l) \; \bar{\nu},  
\end{align*}
 The other fluid kinetic energy term in~(\ref{eq:totenergy}) transforms in a similar way.

Next, the potential energy term in~(\ref{eq:totenergy}) has to be written in a body-fixed frame. For this first write the potential energy term in its original form, 
\begin{align*}
\rho_f g \int_{\mathcal{S}} \int_0^\eta  z \; dz \nu_s \equiv \int_{D \cup B} f \; \mu, 
\end{align*}
where $f: D \cup B \rightarrow \mathbb{R}$ and $\mu= dz \nu_s$. Think of $D \cup B$ as transformed domain from the domain in the body-fixed frame, which is denoted by $\bar{D} \cup \bar{B}$, under the map $\psi_h$. Using the change of variables theorem again, 
\begin{align*}
\int_{D \cup B} f (r) \; \mu&=\int_{\psi_h \left(\bar{D} \cup \bar{B}\right)} f (r) \; \mu, \\
&=\int_{\bar{D} \cup \bar{B}} f (\psi_h(l))\; \bar{\mu}.
\end{align*}
 It is not hard to see that $f \circ \psi_h$  denotes the perpendicular distance from the transformed surface $\bar{\mathcal{S}}$ in the body-fixed frame, and so the potential energy term in the body-fixed frame can be written as
\[\frac{1}{2} \rho_f g\int_{\bar{\mathcal{S}}} \bar{\eta}^2 \; \bar{\nu}_s, \]
where $\bar{\eta}$ is the value of $f \circ \psi_h$ for a point on the free surface. The rest potential energy transforms in a similar way.

The total energy, for the neutrally buoyant case (with $\rho_f=\rho_b=\rho$), referred to the body fixed frame, is therefore  
\begin{align}
K.E. + P.E.
&= \frac{\rho}{2} \left(\int_{\bar{\Sigma}_f} \Phi' \nabla_b \Phi' \cdot \bar{n}_f \; \bar{\nu}  - \int_{\bar{\Sigma}_b} \Phi' \nabla_b \Phi' \cdot \bar{n}_b \; \bar{\nu} \right)   \nonumber \\
& \hspace{1in}+  \frac{1}{2} \rho g \left(\int_{\bar{\mathcal{S}}} \bar{\eta}^2 \; \bar{\nu}_s - \int_{\bar{\mathcal{S}}} \bar{\eta}_0^2 \; \bar{\nu}_s\right)\nonumber \\
& \hspace{2in}   + \frac{1}{2}   \left< \left<(\bar{V},\bar{\Omega}), \rho M_b \cdot  (\bar{V},\bar{\Omega}) \right>\right>,  \label{eq:totenergybody}
\end{align}
Now write this using the variables $(\mathfrak{L}, \mathfrak{A})$. To do this, first rewrite~(\ref{eq:bodyfixedL}) and~(\ref{eq:bodyfixedA}) using~(\ref{eq:vecidL}) and~(\ref{eq:vecidA})  as 
\begin{align}
\left(\mathfrak{L}, \mathfrak{A} \right) &=\rho M_b \cdot \left( \bar{V}, \bar{\Omega} \right) + \rho \bar{P}_{f} \label{eq:lvrel}
\end{align} 
where 
\begin{align}
\bar{P}_{f} \equiv \left(\bar{L}_{f}, \bar{A}_{f} \right)&:=  \left( \underbrace{- \int_{\bar{\Sigma}_b} \Phi' \bar{n}_b \; \bar{\nu}}_{\bar{L}_{f1}} +   \underbrace{\int_{\bar{\Sigma}_f} \Phi' \bar{n}_f  \; \bar{\nu}}_{\bar{L}_{f2}}, \underbrace{- \int_{\bar{\Sigma}_b} l \times \Phi' \bar{n}_b \; \bar{\nu}}_{\bar{A}_{f1}} +  \underbrace{\int_{\bar{\Sigma}_f} l  \times  \Phi' \bar{n}_f \; \bar{\nu}}_{\bar{A}_{f2}}  \right), \label{eq:Pf}
\end{align}
Inverting, obtain
\begin{align}
\left( \bar{V}, \bar{\Omega} \right)&= \left(M_b \right)^{-1} \cdot \left[\frac{1}{\rho}\left(\mathfrak{L}, \mathfrak{A} \right)  - \bar{P}_{f} \right], \label{eq:vlrel}
\end{align}

\subsection{The variables and the variations in the body-fixed frame.}
 Consider now the variables 
\[\left(\bar{\Sigma}_f,\phi_f', \mathfrak{L},\mathfrak{A} \right),\]
where 
\[\phi_f':=\Phi'_{\mid_{{\bar{\Sigma}_f}}}\]
and $\bar{\Sigma}_f$, rather than $\bar{\eta}$, is the variable that will be used to characterize the free surface. As in \cite{LeMaMoRa1986} and \cite{Sh2016}, view $\bar{\Sigma}_f$ as the image of a smooth embedding of a reference configuration of the free surface $\Sigma_0$ which could, without loss of generality, be taken as the undisturbed surface. 
Note that, as in \cite{Sh2016}, the variations in $\bar{\Sigma}_f$ are those that are normal to the fluid surface, and will be denoted either by the vector $\left(\delta \mathbf{\bar{\Sigma}}_f \right)_n$ or its magnitude $\left(\delta \bar{\Sigma}_f \right)_n:=\left(\delta \mathbf{\bar{\Sigma}}_f \right)_n \cdot \bar{n}_f$. These variations are related to $\delta \bar{\eta}$ by 
\begin{align}
\left(\delta \bar{\Sigma}_f \right)_n&=\delta \bar{\eta} \left(\bar{k} \cdot \bar{n}_f\right),  \nonumber \\
\Rightarrow \left(\delta \bar{\Sigma}_f \right)_n \bar{\nu}&=\delta \bar{\eta} \; \bar{\nu}_s, \label{eq:sigeta}
\end{align}
where $\bar{k}$ is the unit vector $k$ in the body-fixed frame.

  In the body-fixed frame, {\it the mixed Dirichlet-Neumann problem} of~(\ref{eq:phi}) takes the form
\begin{align}
\nabla_b^2 \Phi' &=0 \: {\rm in} \: \bar{D} , \quad \Phi' \mid_{\bar{\Sigma}_f} {\rm prescribed},  \: \nabla_b \Phi' \cdot \bar{n}_b \mid_{\bar{\Sigma}_b}= \bar{U} \cdot \bar{n}_b, \nonumber \\
& \hspace{1in}  \Phi' \rightarrow 0 \; {\rm as} \; x_l,y_l \rightarrow \pm \infty, z_l \rightarrow - \infty  \label{eq:phibf} 
\end{align}
where $(x_l,y_l,z_l)=\psi_h^{-1} \left(x,y,z \right)$.

 Examining relations~(\ref{eq:lvrel}) and~(\ref{eq:Pf}) it is seen that, due to the mixed Dirichlet-Neumann problem, $\bar{P}_f$ is not independent of the rigid body's velocities $(\bar{V},\bar{\Omega})$. Otherwise, variations in $\left(\bar{\Sigma}_f,\phi_f'\right)$ could be performed keeping $\left(\mathfrak{L},\mathfrak{A} \right)$ constant and vice-versa, by making appropriate variations in $\left(\bar{V},\bar{\Omega}\right)$. Indeed such is the case in the problem of a rigid body dynamically interacting with singular vortices, {\it op. cit.}

The variations therefore need to be performed more carefully and this warrants a discussion.

Consider the following linear maps. First,
\begin{align*}
L_K: \mathbb{R}^6 \rightarrow C^\infty \left(\mathbb{R}^{3}, \mathbb{R} \right) 
\end{align*}
This is the linear map associated with the {\it Kirchhoff problem} in $\mathbb{R}^{3}$:
\begin{align}
\nabla_b^2 \Phi' &=0 \: {\rm in} \: \mathbb{R}^{3} \backslash B, \quad \nabla_b \Phi' \cdot \bar{n}_b \mid_{\bar{\Sigma}_b}= \bar{U} \cdot \bar{n}_b, \nonumber \\
& \hspace{2in}  \Phi' \rightarrow {\rm 0 \; as \;} l \rightarrow \infty \label{eq:kirchhoff} 
\end{align}
As is well-known in the Kirchhoff problem \cite{Kirchhoff1869, Mi1996}, 
\begin{align}
\Phi'(l,t)=\left(\Psi'(l), \zeta'(l) \right)\cdot  (\bar{V}(t),\bar{\Omega}(t)), \label{eq:decompos}
\end{align}
 where
\begin{align}
\Psi'(l) & \equiv \left( \Psi_x'(l), \Psi_y'(l), \Psi_z'(l)\right), \zeta'(l) \equiv \left( \zeta_x'(l), \zeta_y'(l), \zeta_z'(l)\right)) \label{eq:components}
\end{align}  are 3-vectors each of whose components satisfy the following Neumann problems
\begin{align}
\nabla_b^2 \Psi'_x &=0 \: {\rm in} \: \bar{D} , \quad \nabla_b \Psi'_x \cdot \bar{n}_b \mid_{\bar{\Sigma}_b}= i \cdot \bar{n}_b, \: \Psi'_x \mid_{\infty}={\rm constant} \label{eq:bcLcomponent}\\
\nabla_b^2 \zeta'_x &=0 \: {\rm in} \: \bar{D} , \quad \nabla_b \zeta'_x \cdot \bar{n}_b \mid_{\bar{\Sigma}_b}= \left( i \times l \right) \cdot \bar{n}_b, \: \zeta'_x \mid_{\infty}={\rm constant} \label{eq:bcAcomponent}
\end{align}
and similarly in the $y_l$- and $z_l$-directions (of the body-fixed frame).\footnote{To avoid notation clutter, $\Psi'_x$ etc. is used instead of $\Psi'_{x_l}$ etc.}

 Next, consider the linear map 
\begin{align*}
L_S: C^\infty \left(\bar{\Sigma}_f, \mathbb{R} \right) \rightarrow C^\infty \left(\bar{D}, \mathbb{R} \right),
\end{align*}
associated with the following mixed Dirichlet-Neumann problem for a stationary body: 
\begin{align}
\nabla_b^2 \Phi' &=0 \: {\rm in} \: \bar{D} , \quad \Phi' \mid_{\bar{\Sigma} _f} {\rm prescribed}, \quad \nabla_b \Phi' \cdot \bar{n}_b \mid_{\bar{\Sigma}_b}= 0, \nonumber \\ 
& \hspace{1in}  \Phi' \rightarrow 0 \; {\rm as} \; x_l,y_l \rightarrow \pm \infty, z_l \rightarrow - \infty \label{eq:stationary} 
\end{align}
Each of these linear maps further gives rise to other linear maps by restricting to the boundaries of $\bar{D}$: 
\begin{align}
&L_{K,b}: \mathbb{R}^6 \rightarrow C^\infty \left(\bar{\Sigma}_b, \mathbb{R} \right),  \label{eq:lkb} \\
&L_{K,f}: \mathbb{R}^6 \rightarrow C^\infty \left(\bar{\Sigma}_f, \mathbb{R} \right),   \label{eq:lkf} \\
&L_{S,b}: C^\infty \left(\bar{\Sigma}_f, \mathbb{R} \right) \rightarrow C^\infty \left(\bar{\Sigma}_b, \mathbb{R} \right) \label{eq:lsb} 
\end{align} 
Finally, consider the linear maps
\begin{align}
\mathcal{I}_B: 
C^\infty\left(\bar{\Sigma}_b,\mathbb{R} \right) \rightarrow \mathbb{R}^6, \quad \mathcal{I}_F: C^\infty\left(\bar{\Sigma}_f,\mathbb{R} \right) \rightarrow \mathbb{R}^6, \label{eq:intmaps}
\end{align}
defined by the integrals $\left(\int_{\bar{\Sigma}_b} \Phi' \bar{n}_b \; \bar{\nu}, \int_{\bar{\Sigma}_b}l \times  \Phi' \bar{n}_b \; \bar{\nu}\right)$ and  $\left(\int_{\bar{\Sigma}_f} \Phi' \bar{n}_f \; \bar{\nu}, \int_{\bar{\Sigma}_f} l \times \Phi' \bar{n}_f \; \bar{\nu}\right)$, respectively. Restricting to the `linear' and `angular' components, respectively, each of these maps can also be identified with a pair of maps: $\mathcal{I}_B \equiv \left(\mathcal{I}_{B1}, \mathcal{I}_{B2}\right)$ and $\mathcal{I}_F \equiv \left(\mathcal{I}_{F1}, \mathcal{I}_{F2}\right)$.

   With these maps in place, the arbitrary and independent variation of each variable in the set $\left(\bar{\Sigma}_f,\phi_f', \mathfrak{L},\mathfrak{A} \right)$  will now be discussed. 

\begin{itemize}
\item[1.] An arbitrary variation $\delta_o \left(\mathfrak{L}, \mathfrak{A} \right)$, with $\delta \phi_f'=\left(\delta \bar{\Sigma}_f \right)_n$=0. Here $\delta_o$ denotes that only one in the pair $\left(\mathfrak{L}, \mathfrak{A} \right)$ is varied while the other is kept fixed.  To achieve this requires an appropriate variation $\delta_o(\bar{V},\bar{\Omega})$ (for otherwise, $\Phi'$ in the domain $\bar{D}$ remains unchanged and hence also $\bar{P}_f$, making it impossible, by~(\ref{eq:lvrel}), to achieve the variation $\delta_o \left(\mathfrak{L}, \mathfrak{A} \right)$). But this induces a variation $\delta \Phi'$ in $\bar{D}$, including at the boundaries, given by 
\begin{align*}
\delta \Phi'_{\mid_{\bar{\Sigma}_b}}&=\delta \Phi'_K + \delta \Phi'_S, \\
\delta \Phi'_{\mid_{\bar{\Sigma}_f}}&=\delta \phi_f'=0
\end{align*}
where 
\begin{align*}
\delta \Phi'_K&= L_{K,b} \left( \delta_o(\bar{V},\bar{\Omega})\right), \\
\delta \Phi'_S&=L_{S,b} \circ \left(- L_{K,f} \left( \delta_o(\bar{V},\bar{\Omega})\right) \right), 
\end{align*}
the minus sign ensuring that the constraint $\delta \phi_f'=0$ is respected. Generally therefore, one has an induced variation $\delta \bar{P}_f$, which is given by 
\begin{align*}
\delta \bar{P}_f&= \delta_o \left( \bar{L}_{f1}, \bar{A}_{f1}\right)= \mathcal{I}_B \circ  \left(L_{K,b} - L_{S,b} \circ L_{K,f} \right)\left( \delta_o(\bar{V},\bar{\Omega})\right),
\end{align*}
the map $\mathcal{I}_B$ acting through one of its components $\mathcal{I}_{B1}$ or $\mathcal{I}_{B2}$. The arbitrary variation $\delta_o \left(\mathfrak{L}, \mathfrak{A} \right)$ is therefore possible only if the variation $\delta_o(\bar{V},\bar{\Omega})$ is chosen such that the equation 
\begin{align}
\delta_o \left(\mathfrak{L}, \mathfrak{A} \right) &=\rho M_b \cdot \delta_o \left( \bar{V}, \bar{\Omega} \right) + \rho \delta_o \left( \bar{L}_{f1}, \bar{A}_{f1}\right) \nonumber  \\
&=\rho \left( M_b +  \mathcal{I}_B \circ  \left(L_{K,b} - L_{S,b} \circ L_{K,f} \right) \right) \cdot \delta_o \left( \bar{V}, \bar{\Omega} \right) \label{eq:case1}
\end{align} is satisfied. To show that it is possible to choose such a $\delta_o(\bar{V},\bar{\Omega})$, it is necessary and sufficient that the linear map $M_b +  \mathcal{I}_B \circ  \left(L_{K,b} - L_{S,b} \circ L_{K,f} \right)$ is invertible.

\item[2.] Next, an arbitrary variation $\delta \phi_f'$, with $\delta \left(\mathfrak{L}, \mathfrak{A} \right)=\left(\delta \bar{\Sigma}_f \right)_n $=0. The variations induced are similar to case 1, but there is an extra term due to the imposed variation $\delta \phi_f'$. The boundary variations are therefore given by 
\begin{align}
\delta \Phi'_{\mid_{\bar{\Sigma}_b}}&=\delta \Phi'_K + \delta \Phi'_S, \nonumber \\
\delta \Phi'_{\mid_{\bar{\Sigma}_f}}&=\delta \phi_f' \label{eq:delphif}
\end{align}where 
\begin{align*}
\delta \Phi'_K&= L_{K,b} \left( \delta(\bar{V},\bar{\Omega})\right), \\
\delta \Phi'_S&=L_{S,b} \circ \left(\delta \phi_f'- L_{K,f} \left( \delta(\bar{V},\bar{\Omega})\right) \right), 
\end{align*}The induced variation $\delta \bar{P}_f$ is now given by
\begin{align*}
\delta \bar{P}_f&=\delta \left(\bar{L}_{f}, \bar{A}_{f} \right), \\
&=\delta \left( \bar{L}_{f1}, \bar{A}_{f1}\right)+\delta \left( \bar{L}_{f2}, \bar{A}_{f2}\right) \\
&= \mathcal{I}_B \circ  \left(L_{K,b} - L_{S,b} \circ L_{K,f} \right)\left( \delta(\bar{V},\bar{\Omega})\right) +\mathcal{I}_B \circ  L_{S,b} \left( \delta \phi_f'  \right) +\mathcal{I}_F \left( \delta \phi_f' \right)
\end{align*} In such a case again, a choice of the variation $\delta(\bar{V},\bar{\Omega})$ is required such that   the equation 
\begin{align}
0&=\rho M_b \cdot \delta \left( \bar{V}, \bar{\Omega} \right) + \rho \delta \left( \bar{L}_{f}, \bar{A}_{f}\right) \nonumber \\
\Rightarrow \mathcal{I}_B \circ  L_{S,b} \left( \delta \phi_f'  \right) +\mathcal{I}_F \left( \delta \phi_f' \right)&= \left(M_b +\mathcal{I}_B \circ  \left( L_{S,b} \circ L_{K,f} - L_{K,b} \right)\right)\left( \delta(\bar{V},\bar{\Omega})\right) \label{eq:varvelfixing}
\end{align} is satisfied.  As in case 1, to show such a $\delta(\bar{V},\bar{\Omega})$ exists for any choice of $\delta \phi_f'$
 requires the map $M_b +\mathcal{I}_B \circ  \left( L_{S,b} \circ L_{K,f} - L_{K,b} \right)$ to be invertible.

\item[3.] Finally, an arbitrary variation $\left(\delta \bar{\Sigma}_f \right)_n $, with $\delta \left(\mathfrak{L}, \mathfrak{A} \right)=\delta \phi_f'$=0. The meaning of $\delta \phi_f'$=0 is explained in \cite{Sh2016}. $\phi_f'$ is viewed as a function of the reference configuration. There is thus an induced change $\delta \Phi' \mid_{\bar{\Sigma}_f}$ given by 
\begin{align}
\delta \Phi' \mid_{\bar{\Sigma}_f}&= - \nabla_b \Phi' \mid_{\bar{\Sigma}_f} \cdot  \left(\delta \bar{\Sigma}_f \right)_n \label{eq:phisig}
\end{align}
There is a perturbed fluid domain $\tilde{D}$ in this case and, generally speaking, $O(\epsilon)$-sized subdomains of $\tilde{D}$ could lie outside $\bar{D}$. Considerations of these subdomains is, however, not necessary to compute the variational derivative with respect to $\left(\delta \bar{\Sigma}_f \right)_n $.

      This case is, therefore, treated just like case 2 on the unperturbed domain, with equation~(\ref{eq:delphif}) replaced by~(\ref{eq:phisig}). The equation determining the choice of $\delta (\bar{V},\bar{\Omega})$ is given by~(\ref{eq:varvelfixing}) with $\delta \phi_f'$ replaced by the $\delta \Phi' \mid_{\bar{\Sigma}_f}$of equation~(\ref{eq:phisig}).

\end{itemize}
The map 
\[\mathcal{I}_B \circ  \left(L_{K,b} - L_{S,b} \circ L_{K,f} \right)\]
which appears in all three cases above, is now examined. 

    First, it should be obvious from the definitions of the maps~(\ref{eq:decompos}),~(\ref{eq:components}),~(\ref{eq:lkb}) and~(\ref{eq:intmaps}) that $\mathcal{I}_B \circ  L_{K,b}: \mathbb{R}^6 \rightarrow \mathbb{R}^6$ is nothing but the symmetric added mass matrix: 
\begin{footnotesize}
\begin{align*}
& M_a \\ \\
& \hspace{-1in} :=\left(\begin{array}{cccccc} \int_{\bar{\Sigma}_b} \Psi_x' n_x \; \bar{\nu} & \int_{\bar{\Sigma}_b} \Psi_y' n_x \; \bar{\nu} & \int_{\bar{\Sigma}_b} \Psi_z' n_x \; \bar{\nu} & \int_{\bar{\Sigma}_b}  \zeta_x' n_x \; \bar{\nu} & \int_{\bar{\Sigma}_b} \zeta_y' n_x \; \bar{\nu} & \int_{\bar{\Sigma}_b} \zeta_z' n_x \; \bar{\nu}  \\ 
\int_{\bar{\Sigma}_b} \Psi_x' n_y \; \bar{\nu} & \int_{\bar{\Sigma}_b} \Psi_y' n_y \; \bar{\nu} & \int_{\bar{\Sigma}_b} \Psi_z' n_y \; \bar{\nu} & \int_{\bar{\Sigma}_b}  \zeta_x' n_y \; \bar{\nu} & \int_{\bar{\Sigma}_b} \zeta_y' n_y \; \bar{\nu} & \int_{\bar{\Sigma}_b} \zeta_z' n_y \; \bar{\nu} \\
\int_{\bar{\Sigma}_b} \Psi_x' n_z \; \bar{\nu} & \int_{\bar{\Sigma}_b} \Psi_y' n_z \; \bar{\nu} & \int_{\bar{\Sigma}_b} \Psi_z' n_z \; \bar{\nu} & \int_{\bar{\Sigma}_b} \zeta_x' n_z  \; \bar{\nu} & \int_{\bar{\Sigma}_b} \zeta_y' n_z \; \bar{\nu} & \int_{\bar{\Sigma}_b} \zeta_z' n_z \; \bar{\nu} \\
 \int_{\bar{\Sigma}_b} \Psi_x' (l \times \bar{n}_b)_ x \; \bar{\nu} & \int_{\bar{\Sigma}_b} \Psi_y' (l \times \bar{n}_b)_ x \; \bar{\nu} & \int_{\bar{\Sigma}_b} \Psi_z' (l \times \bar{n}_b)_ x \; \bar{\nu} & \int_{\bar{\Sigma}_b}  \zeta_x' (l \times \bar{n}_b)_ x \; \bar{\nu} & \int_{\bar{\Sigma}_b} \zeta_y' (l \times \bar{n}_b)_ x  \; \bar{\nu} & \int_{\bar{\Sigma}_b} \zeta_z' (l \times \bar{n}_b)_ x \; \bar{\nu} \\ 
\int_{\bar{\Sigma}_b} \Psi_x' (l \times \bar{n}_b)_ y \; \bar{\nu} & \int_{\bar{\Sigma}_b} \Psi_y' (l \times \bar{n}_b)_ y \; \bar{\nu} & \int_{\bar{\Sigma}_b} \Psi_z' (l \times \bar{n}_b)_ y \; \bar{\nu} & \int_{\bar{\Sigma}_b}  \zeta_x' (l \times \bar{n}_b)_ y \; \bar{\nu} & \int_{\bar{\Sigma}_b} \zeta_y' (l \times \bar{n}_b)_ y  \; \bar{\nu} & \int_{\bar{\Sigma}_b} \zeta_z' (l \times \bar{n}_b)_ y \; \bar{\nu} \\
\int_{\bar{\Sigma}_b} \Psi_x' (l \times \bar{n}_b)_ z \; \bar{\nu} & \int_{\bar{\Sigma}_b} \Psi_y' (l \times \bar{n}_b)_ z \; \bar{\nu} & \int_{\bar{\Sigma}_b} \Psi_z' (l \times \bar{n}_b)_ z \; \bar{\nu} & \int_{\bar{\Sigma}_b}  \zeta_x' (l \times \bar{n}_b)_ z \; \bar{\nu} & \int_{\bar{\Sigma}_b} \zeta_y' (l \times \bar{n}_b)_ z  \; \bar{\nu} & \int_{\bar{\Sigma}_b} \zeta_z' (l \times \bar{n}_b)_ z \; \bar{\nu}  \end{array} \right)
\end{align*}
\end{footnotesize}
 Recall, that the symmetry is shown using the boundary conditions in ~(\ref{eq:bcLcomponent}) and~(\ref{eq:bcAcomponent}) and invoking the following well-known reciprocity result for any two harmonic functions  $f$ and $g$ in $\mathbb{R}^{3}$ satisfying the Kirchhoff problem:  
\begin{align}
\int_{\bar{\Sigma}_b} \left(f \nabla_b g \cdot  \bar{n}_b - g \nabla_b f \cdot  \bar{n}_b  \right) \; \bar{\nu}&= \int_{\mathbb{R}^{3}} \left( f \nabla_b^2 g  - g \nabla_b^2 f \right) \; \mu, \label{eq:identity} \\
\Rightarrow \int_{\bar{\Sigma}_b} \left(f \nabla_b g \cdot  \bar{n}_b - g \nabla_b f \cdot  \bar{n}_b   \right) \; \bar{\nu} &=0 \nonumber
\end{align}

 Next, consider the map $-\mathcal{I}_B \circ  \left( L_{S,b} \circ L_{K,f} \right): \mathbb{R}^6 \rightarrow \mathbb{R}^6$. Referring to~(\ref{eq:decompos}),~(\ref{eq:components}),~(\ref{eq:stationary}),~(\ref{eq:lkb}) and~(\ref{eq:intmaps}), this map is given by a coupling matrix, denoted by $M_c$, whose elements are the elements of $M_a$ replaced in the following manner:
\begin{align*}
\int_{\bar{\Sigma}_b} \Psi_x' n_x \; \bar{\nu} & \rightarrow -\int_{\bar{\Sigma}_b} L_{S,b} \left(\Psi_x'   \mid_{\bar{\Sigma}_f}\right)n_x\; \bar{\nu}, \\ \int_{\bar{\Sigma}_b} \Psi_x' (l \times \bar{n}_b)_ x  \; \bar{\nu} &\rightarrow - \int_{\bar{\Sigma}_b}  L_{S,b} \left( \Psi_x' \mid_{\bar{\Sigma}_f}\right)(l \times \bar{n}_b)_ x  \; \bar{\nu},
\end{align*}
etc.  Now from~(\ref{eq:stationary}), 
\[\int_{\bar{\Sigma}_b} \nabla_b \left( L_{S} \left(\Psi_x'   \mid_{\bar{\Sigma}_f}\right) \right) \cdot \bar{n}_b \; \bar{\nu}=0,\]
etc. Use this boundary condition in the identity~(\ref{eq:identity}), with 
\[f:=\Psi_x' + L_{S} \left(\Psi_x'   \mid_{\bar{\Sigma}_f}\right),\quad g:=\Psi_y' + L_{S} \left(\Psi_y'   \mid_{\bar{\Sigma}_f}\right)\]
Since $\Psi_x'$and $\Psi_y'$ already satisfy the reciprocity result, one obtains: 
\begin{align*}
\int_{\bar{\Sigma}_b} \left(  L_{S,b} \left(\Psi_x'   \mid_{\bar{\Sigma}_f}\right)    \nabla_b \Psi_y' \cdot  \bar{n}_b -  L_{S,b} \left(\Psi_y'   \mid_{\bar{\Sigma}_f}\right) \nabla_b \Psi_x' \cdot  \bar{n}_b   \right) \; \bar{\nu} &=0, \: {\rm etc.}
\end{align*}
Using~(\ref{eq:bcLcomponent}) and~(\ref{eq:bcAcomponent}) again, this shows that $M_c$ is also a symmetric matrix. 

       Therefore, the arbitrary and independent variations discussed previously are possible if and only if the $6 \times 6$ symmetric matrix 
\begin{align}
M&:=M_b+ M_a+M_c, \label{eq:matrixM}
\end{align}
is invertible. Note that case 1 requires only the invertibility of the upper left or lower right $3 \times 3$ blocks of $M$. The invertibility of $M$ is not examined in this paper, and it is assumed to be invertible. 

\subsection{Phase space and Hamiltonian formalism.}

 Consider the space of $\left(\bar{\Sigma}_f,\phi_f', \mathfrak{L},\mathfrak{A} \right)$, denoted by 
 \[\mathcal{Z}_b \times \mathfrak{se}(3)^*.\] On this space define the Hamiltonian function as the total energy function~(\ref{eq:totenergybody}) written in terms of these variables:
\begin{align}
 & H \left(\bar{\Sigma}_f,\phi_f', \mathfrak{L},\mathfrak{A} \right) \nonumber \\
&=\frac{\rho}{2} \left(\int_{\bar{\Sigma}_f} \Phi' \nabla_b \Phi' \cdot \bar{n}_f \; \bar{\nu}  - \int_{\bar{\Sigma}_b} \Phi' \nabla_b \Phi' \cdot \bar{n}_b \; \bar{\nu} \right) + \frac{1}{2} \rho g\int_{\bar{\mathcal{S}}} \bar{\eta}^2 \; \bar{\nu}_s \nonumber \\
& \hspace{1in}   + \frac{1}{2}   \left< \left< \left(M_b \right)^{-1} \cdot \left[\frac{1}{\rho}\left(\mathfrak{L}, \mathfrak{A} \right)  - \bar{P}_{f} \right], \rho M_b \cdot  \left(M_b \right)^{-1} \cdot \left[\frac{1}{\rho}\left(\mathfrak{L}, \mathfrak{A} \right)  - \bar{P}_{f} \right] \right>\right>,  \nonumber \\ \nonumber 
&= \frac{\rho}{2} \left(\int_{\bar{\Sigma}_f} \Phi' \nabla_b \Phi' \cdot \bar{n}_f \; \bar{\nu}  - \int_{\bar{\Sigma}_b} \Phi' \nabla_b \Phi' \cdot \bar{n}_b \; \bar{\nu} \right) + \frac{1}{2} \rho g\int_{\bar{\mathcal{S}}} \bar{\eta}^2 \; \bar{\nu}_s \nonumber \\
& \hspace{1in}   + \frac{1}{2}   \left< \left< \left(M_b \right)^{-1} \cdot \left[\frac{1}{\rho}\left(\mathfrak{L}, \mathfrak{A} \right)  - \bar{P}_{f} \right], \rho  \left[\frac{1}{\rho}\left(\mathfrak{L}, \mathfrak{A} \right)  - \bar{P}_{f} \right] \right>\right>,   \label{eq:hamfunc}
\end{align}

  Now consider the following Poisson brackets on $\mathcal{Z}_b \times \mathfrak{se}(3)^*$, 
\begin{align}
\{F,G \}&:= \{F_{\mid_{\mathcal{Z}_b}},G_{\mid_{\mathcal{Z}_b}}\}_{{\rm Zakharov}}+ \{F_{\mid_{\mathfrak{se}(3)^*}},G_{\mid_{\mathfrak{se}(3)^*}} \}_{{\rm Lie-Poisson}} \label{eq:pb}
\end{align}
where $\{F_{\mid_{\mathcal{Z}_b}},G_{\mid_{\mathcal{Z}_b}}\}_{{\rm Zakharov}}$  is the Zakharov bracket but written in the variables $\left(\bar{\Sigma}_f,\phi_f' \right)$,  and given by
\begin{align*}
\{\hat{F},\hat{G}\}_{{\rm Zakharov}}&:=\frac{1}{\rho} \int_{\bar{\Sigma}_f} \left( \frac{\delta \hat{f}}{ \left(\delta \bar{\Sigma}_f \right)_n} \frac{\delta \hat{g}}{\delta \phi_f'} - \frac{\delta \hat{g}}{ \left(\delta \bar{\Sigma}_f \right)_n} \frac{\delta \hat{f}}{\delta \phi_f'} \right) \; \bar{\nu}, 
\end{align*}
for functions $\hat{F},\hat{G}:\mathcal{Z}_b \rightarrow \mathbb{R} $ of the form 
\begin{align*}
\hat{F}&=\int_{\bar{\Sigma}_f} \hat{f} \; \bar{\nu},
\end{align*}
etc. And $\{F_{\mid_{\mathfrak{se}(3)^*}},G_{\mid_{\mathfrak{se}(3)^*}} \}_{{\rm Lie-Poisson}}$ is the negative Lie-Poisson bracket on $\mathfrak{se}(3)^* \equiv  \mathbb{R}^{3*} \times  \mathbb{R}^{3*} $, given by 
\begin{align}
  \{\tilde{F}, \tilde{G} \}_{{\rm Lie-Poisson}} &:=- \left\langle \mu,\left[\frac{\partial \tilde{F}}{\partial \mu},\frac{\partial \tilde{G}}{\partial \mu}\right]\right\rangle, \nonumber \\
&= \left\langle \frac{\partial \tilde{F} } {\partial \mu}, \operatorname{ad}^{*}_{\partial \tilde{G}/ \partial \mu} \mu
\right\rangle, \label{eq:liepoiss}
\end{align}
for $ \tilde{F}, \tilde{G}: \mathfrak{se}(3)^*  \rightarrow \mathbb{R}$ and $\mu \in \mathfrak{se}(3)^* \equiv \mathbb{R}^6$ \cite{MaRa99}.

\paragraph{Functional Derivatives.} Compute now the various functional derivatives of $H$ corresponding to the variations 1, 2 and 3, described previously.  

Starting with case 1, the variation in the Hamiltonian is computed as 
\begin{align*}
& \left< \left< \frac{\delta H}{\delta \left(\mathfrak{L},\mathfrak{A} \right)}, \delta \left(\mathfrak{L},\mathfrak{A} \right)\right> \right> \\
&= \frac{1}{\epsilon} \lim_{\epsilon \rightarrow 0}  \left[H \left(\bar{\Sigma}_f,\phi_f', \left(\mathfrak{L},\mathfrak{A} \right) + \epsilon \delta(\mathfrak{L},\mathfrak{A})  \right) - H \left(\bar{\Sigma}_f,\phi_f', \mathfrak{L},\mathfrak{A} \right)\right] \\
&= \frac{1}{\epsilon} \lim_{\epsilon \rightarrow 0} \left[\frac{\epsilon \rho}{2}\left(\int_{\bar{\Sigma}_f} \Phi' \nabla_b \delta \Phi' \cdot \bar{n}_f \; \bar{\nu}- \int_{\bar{\Sigma}_b} \delta \Phi' \nabla_b  \Phi' \cdot \bar{n}_b \; \bar{\nu}_b - \int_{\bar{\Sigma}_b}\Phi' \nabla_b  \delta \Phi' \cdot \bar{n}_b \; \bar{\nu}_b\right) \right. \\ \\
& +  \frac{\epsilon}{2}   \left< \left< \left(M_b \right)^{-1} \cdot \left[\frac{1}{\rho} \delta \left(\mathfrak{L}, \mathfrak{A} \right)  - \delta \left(\bar{L}_{f1}, \bar{A}_{f1} \right) \right], \rho  \left[\frac{1}{\rho}\left(\mathfrak{L}, \mathfrak{A} \right)  - \bar{P}_{f} \right] \right>\right>  \\ \\
&+ \left. \frac{\epsilon}{2} \left< \left< \left(M_b \right)^{-1} \cdot \left[\frac{1}{\rho}  \left(\mathfrak{L}, \mathfrak{A} \right)  -  \left(\bar{L}_{f1}, \bar{A}_{f1} \right) \right], \rho  \left[\frac{1}{\rho} \delta \left(\mathfrak{L}, \mathfrak{A} \right)  - \delta \left(\bar{L}_{f1}, \bar{A}_{f1} \right)\right] \right>\right>  \right],
\end{align*}
where $\delta \Phi'$ is the induced variation in this case, as discussed previously. In the first integral on the right, it should be noted that though $\delta \Phi' \mid_{\bar{\Sigma}_f}$=0 (since $\delta \phi_f'=\delta \bar{\Sigma}_f$=0), $\nabla_b \delta \Phi'$could be non-zero. 
Now use the well-known identity for two harmonic functions $f,g$ in a domain with boundaries
\begin{align*} 
\int_{\partial D} f \nabla g \cdot n \; \nu-\int_{\partial D} g \nabla f \cdot n \; \nu&=0
\end{align*} 
 Apply this to the functions $\Phi'$ and $\delta \Phi'$ and with $\partial D \equiv \bar{\Sigma}_f \cup \bar{\Sigma}_b \cup \mathcal{S}$. The normal derivatives of both the functions vanish at $\mathcal{S}$, leading to the relation 
\begin{align}
\int_{\bar{\Sigma}_f} \Phi' \nabla_b \delta \Phi' \cdot \bar{n}_f \; \bar{\nu} - \int_{\bar{\Sigma}_b} \Phi' \nabla_b \delta \Phi' \cdot \bar{n}_b \; \bar{\nu}_b&=  \int_{\bar{\Sigma}_f} \delta \Phi' \nabla_b \Phi' \cdot \bar{n}_f \; \bar{\nu} - \int_{\bar{\Sigma}_b} \delta \Phi' \nabla_b  \Phi' \cdot \bar{n}_b \; \bar{\nu}_b,  \label{eq:identity} \\
&=- \int_{\bar{\Sigma}_b} \delta \Phi' \nabla_b  \Phi' \cdot \bar{n}_b \; \bar{\nu}_b \nonumber 
\end{align} 
And so  
\begin{align*} 
& \frac{1}{\epsilon} \lim_{\epsilon \rightarrow 0}  \left[H \left(\bar{\Sigma}_f,\phi_f', \mathfrak{L},\mathfrak{A} + \epsilon \delta(\mathfrak{L},\mathfrak{A})  \right) - H \left(\bar{\Sigma}_f,\phi_f', \mathfrak{L},\mathfrak{A} \right)\right] \\
&= \frac{1}{\epsilon} \lim_{\epsilon \rightarrow 0} \left[- \epsilon \rho \int_{\bar{\Sigma}_b} \delta \Phi' \nabla_b  \Phi' \cdot \bar{n}_b \; \bar{\nu}_b + \epsilon \rho \left< \left<\left(\bar{V},\bar{\Omega} \right), \frac{1}{\rho} \delta \left(\mathfrak{L}, \mathfrak{A} \right)  - \delta \left(\bar{L}_{f1}, \bar{A}_{f1} \right) \right> \right> \right], \\
&=\left[- \rho \int_{\bar{\Sigma}_b} \delta \Phi' \left(\bar{V}+ \bar{\Omega} \times l \right) \cdot \bar{n}_b \; \bar{\nu}_b +  \left< \left< \left(\bar{V},\bar{\Omega} \right), \delta \left(\mathfrak{L}, \mathfrak{A} \right)  - \rho \delta \left(\bar{L}_{f1}, \bar{A}_{f1} \right) \right> \right> \right], \\ \\
&=\left[\rho  \left< \left< \left(\bar{V}, \bar{\Omega}\right), \delta \left(\bar{L}_{f1}, \bar{A}_{f1} \right)  \right> \right> +  \left< \left< \left(\bar{V},\bar{\Omega} \right), \delta \left(\mathfrak{L}, \mathfrak{A} \right)  - \rho \delta \left(\bar{L}_{f1}, \bar{A}_{f1} \right) \right> \right> \right],  \\ \\
& \hspace{0.5in} [{\rm from~(\ref{eq:Pf})}] 
\end{align*} 
from which is obtained  
\begin{align*}
\frac{\partial H}{\partial (\mathfrak{A}, \mathfrak{L})}&= \left( \bar{\Omega}, \bar{V} \right)
\end{align*}

  Consider next a case 2 variation, 
\begin{align*}
& \int_{\bar{\Sigma_f}} \frac{\delta h}{\delta \phi_f'}\delta \phi_f' \; \nu \equiv \int_{\bar{\Sigma_f}} \frac{\delta h}{\delta \phi_f'}\delta \Phi' \; \nu \\
&= \frac{1}{\epsilon} \lim_{\epsilon \rightarrow 0}  \left[H \left(\bar{\Sigma}_f,\phi_f'+\epsilon \delta \phi_f', \mathfrak{L},\mathfrak{A}   \right) - H \left(\bar{\Sigma}_f,\phi_f', \mathfrak{L},\mathfrak{A} \right)\right] \\
&= \frac{1}{\epsilon} \lim_{\epsilon \rightarrow 0} \left[\frac{\epsilon \rho}{2}\left(\int_{\bar{\Sigma}_f} \delta \Phi' \nabla_b \Phi' \cdot \bar{n}_f \; \bar{\nu}+\int_{\bar{\Sigma}_f} \Phi' \nabla_b \delta \Phi' \cdot \bar{n}_f \; \bar{\nu}- \int_{\bar{\Sigma}_b} \delta \Phi' \nabla_b  \Phi' \cdot \bar{n}_b \; \bar{\nu}_b - \int_{\bar{\Sigma}_b}\Phi' \nabla_b  \delta \Phi' \cdot \bar{n}_b \; \bar{\nu}_b\right) \right. \\ \\
& +  \frac{\epsilon}{2}   \left< \left< \left(M_b \right)^{-1} \cdot \left[ - \delta \left(\bar{L}_{f}, \bar{A}_{f} \right) \right], \rho  \left[\frac{1}{\rho}\left(\mathfrak{L}, \mathfrak{A} \right)  - \bar{P}_{f} \right] \right>\right>  \\ \\
&+ \left. \frac{\epsilon}{2} \left< \left< \left(M_b \right)^{-1} \cdot \left[\frac{1}{\rho}  \left(\mathfrak{L}, \mathfrak{A} \right)  -  \left(\bar{L}_{f}, \bar{A}_{f} \right) \right], \rho  \left[  - \delta \left(\bar{L}_{f}, \bar{A}_{f} \right)\right] \right>\right>  \right], \\ \\
&= \frac{1}{\epsilon} \lim_{\epsilon \rightarrow 0} \left[\epsilon \rho\left(\int_{\bar{\Sigma}_f} \delta \Phi' \nabla_b \Phi' \cdot \bar{n}_f \; \bar{\nu}- \int_{\bar{\Sigma}_b} \delta \Phi' \nabla_b  \Phi' \cdot \bar{n}_b \; \bar{\nu}_b \right) +   \epsilon  \left< \left< \left[ - \delta \left(\bar{L}_{f}, \bar{A}_{f} \right) \right], \rho \left(\bar{V}, \bar{\Omega}  \right) \right>\right>   \right], \\ \\
&[{\rm using~(\ref{eq:identity})}]  \\ \\
&= \rho\left(\int_{\bar{\Sigma}_f} \delta \Phi' \nabla_b \Phi' \cdot \bar{n}_f \; \bar{\nu}  +  \left< \left< \left[ - \delta \left(\bar{L}_{f2}, \bar{A}_{f2} \right) \right],  \left(\bar{V}, \bar{\Omega}  \right) \right>\right> \right),  \\ \\
&= \rho\left(\int_{\bar{\Sigma}_f} \delta \Phi' \nabla_b \Phi' \cdot \bar{n}_f \; \bar{\nu}  +  \left< \left< \left[ -  \left(\int_{\bar{\Sigma}_f} \delta \Phi' \bar{n}_f  \; \bar{\nu},\int_{\bar{\Sigma}_f} l  \times  \delta \Phi' \bar{n}_f \; \bar{\nu} \right) \right],  \left(\bar{V}, \bar{\Omega}  \right) \right>\right> \right), \\ \\
&=\rho\left(\int_{\bar{\Sigma}_f} \delta \Phi' \left( \nabla_b \Phi' - \bar{V} - \bar{\Omega} \times l  \right)\cdot \bar{n}_f \; \bar{\nu}  \right), 
\end{align*}
which implies that 
\begin{align*}
\frac{\delta h}{\delta \phi_f'}&= \rho \left( \nabla_b \Phi' - \bar{V} - \bar{\Omega} \times l  \right)\cdot \bar{n}_f 
\end{align*}

 Finally, consider variation case 3. Note that to be consistent with~(\ref{eq:mass}), these variations must satisfy 
\begin{align}
\rho g \int_{\bar{\mathcal{S}}} \delta \bar{\eta} \; \bar{\nu}_s&=0, \label{eq:varmass}
\end{align}
\begin{align*}
& \int_{\bar{\Sigma_f}} \frac{\delta h}{\left(\delta \bar{\Sigma}_f \right)_n} \left(\delta \bar{\Sigma}_f \right)_n \; \bar{\nu} \\
&= \frac{1}{\epsilon} \lim_{\epsilon \rightarrow 0}  \left[H \left(\bar{\Sigma}_f+\epsilon \delta \bar{\Sigma}_f ,\phi_f', \mathfrak{L},\mathfrak{A}   \right) - H \left(\bar{\Sigma}_f,\phi_f', \mathfrak{L},\mathfrak{A} \right)\right] \\
&= \frac{1}{\epsilon} \lim_{\epsilon \rightarrow 0} \left[\frac{\rho}{2} \left(\delta \int_{\bar{\Sigma}_f}  \Phi' \nabla_b \Phi' \cdot \bar{n}_f \; \bar{\nu}- \epsilon \left(\int_{\bar{\Sigma}_b} \delta \Phi' \nabla_b  \Phi' \cdot \bar{n}_b \; \bar{\nu} + \int_{\bar{\Sigma}_b}\Phi' \nabla_b  \delta \Phi' \cdot \bar{n}_b \; \bar{\nu} \right) \right) \right. \\ \\
& +  \frac{\epsilon}{2}   \left< \left< \left(M_b \right)^{-1} \cdot \left[ - \delta \left(\bar{L}_{f}, \bar{A}_{f} \right) \right], \rho  \left[\frac{1}{\rho}\left(\mathfrak{L}, \mathfrak{A} \right)  - \bar{P}_{f} \right] \right>\right>  \\ \\
& + \left. \frac{\epsilon}{2} \left< \left< \left(M_b \right)^{-1} \cdot \left[\frac{1}{\rho}  \left(\mathfrak{L}, \mathfrak{A} \right)  -  \left(\bar{L}_{f}, \bar{A}_{f} \right) \right], \rho  \left[  - \delta \left(\bar{L}_{f}, \bar{A}_{f} \right)\right] \right>\right>  \right] \\
& \hspace{3in} +   \rho g \int_{\bar{\mathcal{S}}} \bar{\eta}\; \delta \bar{\eta} \; \bar{\nu}_s , 
\end{align*}
where the $\delta \int_{\bar{\Sigma}_f}  \Phi' \nabla_b \Phi' \cdot \bar{n}_f \; \bar{\nu}$ term is calculated as in the problem without the rigid body \cite{Zakharov1968}; see also \cite{Sh2016}.

 Importing this term, obtain
\begin{align*}
& \int_{\bar{\Sigma_f}} \frac{\delta h}{\left(\delta \bar{\Sigma}_f \right)_n} \left( \delta \bar{\Sigma_f} \right)_n\; \nu \\
&= \frac{1}{\epsilon} \lim_{\epsilon \rightarrow 0}  \left[H \left(\bar{\Sigma}_f+\epsilon \delta \bar{\Sigma}_f ,\phi_f', \mathfrak{L},\mathfrak{A}   \right) - H \left(\bar{\Sigma}_f,\phi_f', \mathfrak{L},\mathfrak{A} \right)\right] \\
&= \frac{1}{\epsilon} \lim_{\epsilon \rightarrow 0} \left[\frac{\rho}{2} \left(\pm \int_{V_\Sigma} \nabla_b \Phi' \cdot \nabla_b \Phi' \; \bar{\mu} +\epsilon \int_{\bar{\Sigma}_f}  \left(   \delta \Phi'  \nabla_b \Phi'  \cdot \bar{n}_f +   \Phi'  \nabla_b \delta \Phi'  \cdot \bar{n}_f   \right)  \; \bar{\nu} \right. \right. \\
&\hspace{1in}  \left. \left. - \epsilon \left(\int_{\bar{\Sigma}_b} \delta \Phi' \nabla_b  \Phi' \cdot \bar{n}_b \; \bar{\nu}_b + \int_{\bar{\Sigma}_b}\Phi' \nabla_b  \delta \Phi' \cdot \bar{n}_b \; \bar{\nu}_b\right) \right) \right. \\ \\
& \hspace{2in} + \left.\epsilon \rho \left< \left< \left(\bar{V}, \bar{\Omega} \right),  \left[  - \delta \left(L_{f}, A_{f} \right)\right] \right>\right> \right.  \\
& \hspace{3in} \left. +   \epsilon \rho g \int_{\bar{\mathcal{S}}} \bar{\eta}\; \left( \delta \bar{\Sigma_f} \right)_n  \; \bar{\nu}  \right],  \\  \\
&[{\rm invoking~(\ref{eq:sigeta}) \; for \; the \; potential \; energy \; term}] \\ \\
&= \frac{1}{\epsilon} \lim_{\epsilon \rightarrow 0} \left[ \pm \frac{\rho}{2} \int_{V_\Sigma} \nabla_b \Phi' \cdot \nabla_b \Phi' \; \bar{\mu} +\epsilon \rho  \int_{\bar{\Sigma}_f}  \delta \Phi'  \nabla_b \Phi'  \cdot \bar{n}_f  \right. \\ \\
& \hspace{2in} + \left.\epsilon \rho \left< \left< \left(\bar{V}, \bar{\Omega} \right),  \left[  - \delta \left(L_{f2}, A_{f2} \right)\right] \right>\right> + \epsilon \rho g \int_{\bar{\mathcal{S}}} \bar{\eta}\; \left( \delta \bar{\Sigma_f} \right)_n \; \bar{\nu}  \right],   \\ \\
&[{\rm using~(\ref{eq:identity}) \; and \; proceeding \; in \; the \; same \; way \; as \; in \; case \; 2} ]  \\ \\
&= \frac{1}{\epsilon} \lim_{\epsilon \rightarrow 0} \left[ \pm \frac{\rho}{2} \int_{V_\Sigma} \nabla_b \Phi' \cdot \nabla_b \Phi' \; \bar{\mu} +\epsilon \rho  \int_{\bar{\Sigma}_f}  \delta \Phi'  \left(\nabla_b \Phi' - \left(\bar{V} +\bar{\Omega} \times l  \right) \right) \cdot \bar{n}_f \right. \\ \\
& \left. \hspace{4in}+ \epsilon \rho g \int_{\bar{\mathcal{S}}} \bar{\eta}\; \left( \delta \bar{\Sigma_f} \right)_n \; \bar{\nu}\right], 
\end{align*}
\begin{align*}
&= \frac{1}{\epsilon} \lim_{\epsilon \rightarrow 0} \left[ \frac{\rho}{2} \epsilon (\delta \bar{\Sigma}_f)_n \int_{\bar{\Sigma}_f}  \nabla_b \Phi' \cdot \nabla_b \Phi' \; \bar{\nu} +\epsilon \rho  \int_{\bar{\Sigma}_f}  \delta \Phi'  \left(\nabla_b \Phi' - \left(\bar{V} +\bar{\Omega} \times l  \right) \right) \cdot \bar{n}_f \right. \\ \\ 
& \left. \hspace{4in} + \epsilon \rho g \int_{\bar{\mathcal{S}}} \bar{\eta}\; \left( \delta \bar{\Sigma_f} \right)_n \; \bar{\nu} \right],  \\
& {\rm [as \; in \; the \; problem \; without \; the \; rigid \; body]} \\
&= \frac{1}{\epsilon} \lim_{\epsilon \rightarrow 0} \left[ \frac{\rho}{2} \epsilon (\delta \bar{\Sigma}_f)_n \int_{\bar{\Sigma}_f}  \nabla_b \Phi' \cdot \nabla_b \Phi' \; \bar{\nu} \right. \\
& \left. \hspace{1in} - \epsilon \rho  \int_{\bar{\Sigma}_f} \nabla_b \Phi'  \cdot \bar{n}_f  \left(\delta \bar{\Sigma}_f \right)_n  \left(\nabla_b \Phi' - \left(\bar{V} +\bar{\Omega} \times l  \right) \right) \cdot \bar{n}_f \right. \\ \\
& \left. \hspace{3.5in} + \epsilon \rho g \int_{\bar{\mathcal{S}}} \left(\bar{\eta}-\bar{\eta}_0 \right)\; \left( \delta \bar{\Sigma_f} \right)_n \; \bar{\nu}\right],  \\
&[{\rm using \;~(\ref{eq:phisig}),~(\ref{eq:varmass}) \; and~(\ref{eq:sigeta})  }]
\end{align*}
And so 
\begin{align*}
\frac{\delta h}{\delta \bar{\Sigma}_f}&= \rho  \left(\frac{1}{2}  \nabla_b \Phi' \cdot \nabla_b \Phi'  -  \left(\nabla_b \Phi'  \cdot \bar{n}_f \right)^2 + \left(\nabla_b \Phi'  \cdot \bar{n}_f \right) \left(\bar{V} +\bar{\Omega} \times l  \right) \cdot \bar{n}_f + g \left(\bar{\eta}-\bar{\eta}_0 \right) \right), 
\end{align*}

Collecting all the functional derivatives, 
\begin{align*}
\frac{\delta h}{\delta \phi_f'}&= \rho \left( \nabla_b \Phi' - \bar{V} - \bar{\Omega} \times l  \right)\cdot \bar{n}_f, \\
\frac{\delta h}{\left(\delta \bar{\Sigma}_f \right)_n}&= \rho  \left(\frac{1}{2}  \nabla_b \Phi' \cdot \nabla_b \Phi'  -  \left(\nabla_b \Phi'  \cdot \bar{n}_f \right)^2 + \left(\nabla_b \Phi'  \cdot \bar{n}_f \right) \left(\bar{V} +\bar{\Omega} \times l  \right) \cdot \bar{n}_f +  g \left(\bar{\eta}-\bar{\eta}_0 \right)\right), \\
\frac{\partial H}{\partial (\mathfrak{A}, \mathfrak{L})}&= \left(\bar{\Omega} , \bar{V}\right)
\end{align*}

The Hamiltonian equations of the motion of the coupled system, with respect to the Poisson brackets~(\ref{eq:pb}), are:
\begin{align}
\frac{\partial \bar{\Sigma}_f}{\partial t}&= \frac{1}{\rho} \frac{\delta h}{\delta \phi_f'}=\left( \nabla_b \Phi' - \bar{V} - \bar{\Omega} \times l  \right)\cdot \bar{n}_f, \label{eq:sigmaeq}\\
\frac{\partial  \phi_f' }{\partial t}&= -\frac{1}{\rho} \frac{\delta h}{ \left(\delta \bar{\Sigma}_f \right)_n} \nonumber \\ \nonumber \\
&= - \left(\frac{1}{2}  \nabla_b \Phi' \cdot \nabla_b \Phi'  -  \left(\nabla_b \Phi'  \cdot \bar{n}_f \right)^2 + \left(\nabla_b \Phi'  \cdot \bar{n}_f \right) \left(\bar{V} +\bar{\Omega} \times l  \right) \cdot \bar{n}_f +  g \left(\bar{\eta}-\bar{\eta}_0 \right)\right), \label{eq:phieq} \\
\frac{d  (\mathfrak{A}, \mathfrak{L})}{dt}&=\operatorname{ad}^{*}_{\partial H / \partial  (\mathfrak{A}, \mathfrak{L})} (\mathfrak{A}, \mathfrak{L})=\left( \mathfrak{A} \times \bar{\Omega} + \mathfrak{L} \times \bar{V}, \mathfrak{L} \times \bar{\Omega}\right) \label{eq:momeq}
\end{align}
It is easily checked that equation~(\ref{eq:momeq}) is the same as equations~(\ref{eq:lmomeq}) and~(\ref{eq:angmomeq}), obtained from the global momentum analysis. Equation~(\ref{eq:phieq}) is Bernoulli's equation at the free surface~(\ref{eq:pf}) in the absence of surface tension ($p=p_{atm}$), after using the following relation \cite{Mi1996, Zakharov1968, Sh2016}
\[\frac{\partial \phi'_f}{\partial t}=\frac{\partial }{\partial t} \left(\Phi'_{\mid_{\bar{\Sigma}_f}} \right)+ \left(\nabla_b \Phi' \cdot \bar{n}_f \right)_{\mid_{\bar{\Sigma}_f}}\frac{\partial \bar{\Sigma}_f}{\partial t} \] 

\section{Summary and future directions.} 
The problem presented in this paper is in a general framework. It would be of particular interest to seek some special configurations, for example,  moving equilibrium configurations involving a rigid body and traveling wave(s), and examine their associated stability. The Hamiltonian formalism would allow a nonlinear stability analysis to be performed, analogous to that done for the F\"{o}ppl equilibrium in the problem of a 2D rigid cylinder and point vortices \cite{ShMaBuKe2002}.  Examining the dynamically coupled interaction of a soliton approaching a neutrally buoyant rigid body would be another interesting direction. 

  From a Hamiltonian and geometric mechanics perspective, it would also be of interest to {\it derive} the Poisson brackets of this paper from well-formulated theories of symmetry and reduction of Hamiltonian systems \cite{MaRa99}, along the lines of \cite{LeMaMoRa1986, BaMaRa2012}. 

  Vortices can be generated by free surfaces, and the problem of the dynamically coupled interaction of a free surface and vortices has also been examined from a Hamiltonian perspective \cite{LeMaMoRa1986, RoWr1993}.  The same is true for the problem of a neutrally buoyant rigid body and vortices \cite{ShMaBuKe2002, Sh2005, ShShKeMa2008}. It would be a natural extension therefore to examine the dynamics interaction problem of a rigid body close to a free surface and in the presence of vortices. Indeed, in the viscous Navier-Stokes setting, this problem for stationary rigid bodies has quite a few interesting features; see, for example,  \cite{BrThLeHo2014} and references therein. 

  Apart from linearization approaches, free surface dynamics has also been studied in various asymptotic limits. The shallow water approximation in particular has proved to be very popular. It would be interesting to see how the presence of a dynamically interacting rigid body could be accommodated in such approximations. Presumably, including parameters based on the body size, could lead to some new asymptotic limits.

\newpage 

\section{Appendix A: Global momentum evolution equations in a spatially-fixed frame.}

 Details of the derivation of equations~(\ref{eq:dLdt}) and~(\ref{eq:dAdt}), valid in a spatially-fixed frame (whose origin is taken at the center of the disc $C_R$),  are presented in this appendix. 

\paragraph{Linear Momentum.} Applying Newton's second law for the evolution of $L_T$ first, at any given time instant $t$,
\begin{align*}
\frac{dL_T}{dt}
&=\lim_{R \rightarrow \infty}\left(- \int_{\Sigma_R} p_{atm}n_f \; \nu +\int_{C_R} p n_s \; \nu - \rho_f g k \int_{C_R} \eta \; \nu - \int_W p e_R \; \nu \right) \\
& \hspace{3in} 
+\rho_fg \mathcal{V}_Bk -\rho_b g \mathcal{V}_Bk,
\end{align*}
where $k$ is unit vector opposite to the gravity direction and coincides with $n_s$. Note that the third
integral on the right in original form is $\int_0^\eta \int_{C_R} dz \; \nu $. 

       Referring to~(\ref{eq:fluidlinmom}) and~(\ref{eq:totmom}), the equation becomes
\begin{align*}
& \frac{d}{dt} \left(\frac{\rho_f}{2}  \int_{\Sigma_b} r \times \left(n_b \times \nabla \Phi \right) \; \nu+L_b \right) \\
& = \lim_{R \rightarrow \infty}  \left( - \rho_f \frac{d}{dt} \int_{\Sigma_R} \Phi n_f  \; \nu -  \int_{\Sigma_R} p_{atm}n_f  \; \nu + \rho_f  \frac{d}{dt} \int_{C_R} \Phi  n_s \; \nu + \int_{C_R} p n_s  \; \nu \right.  \\
& \left. \hspace{1in} - \rho_f g k \int_{C_R} \eta \; \nu - \int_W p e_R \; \nu \right) 
+\rho_fg \mathcal{V}_Bk -\rho_b g \mathcal{V}_Bk , \\
& = \lim_{R \rightarrow \infty}  \left( - \rho_f \frac{d}{dt} \int_{\Sigma_R} \Phi n_f  \; \nu -  \int_{\Sigma_R} p_{atm}n_f  \; \nu + \int_{C_R}\left(\rho_f   \frac{\partial  \Phi}{\partial t}+ p \right) \; n_s \nu \right.  \\
& \left. \hspace{1in} - \rho_f g k \int_{C_R} \eta \; \nu - \int_W p e_R \; \nu \right) 
+g \mathcal{V}_Bk \left(\rho_f -\rho_b  \right) , \\
& [{\rm at \; any \;} R \;  {\rm fixed \; in \; time}] \\
&=\lim_{R \rightarrow \infty}  \left( - \rho_f \frac{d}{dt} \int_{\Sigma_R} \Phi n_f  \; \nu -  \int_{\Sigma_R} p_{atm}n_f  \; \nu + \int_{C_R}\left( p_{atm} + \rho_f g \eta_0 -\rho_f  \frac{\nabla \Phi \cdot \nabla \Phi}{2} \right) \; n_s \nu \right.  \\
& \left. \hspace{1in} - \rho_f g k \int_{C_R} \eta \; \nu - \int_W p e_R \; \nu \right) 
+g \mathcal{V}_Bk \left(\rho_f -\rho_b  \right) , \\
& [{\rm using \;}~(\ref{eq:ps})] \\
&=\lim_{R \rightarrow \infty}  \left( - \rho_f \frac{d}{dt} \int_{\Sigma_R} \Phi n_f  \; \nu -  \int_{\Sigma_R} p_{atm}n_f  \; \nu + \int_{C_R}\left( p_{atm} -\rho_f  \frac{\nabla \Phi \cdot \nabla \Phi}{2} \right) \; n_s \nu \right.  \\
& \left. \hspace{1in} + \rho_f g k \int_{C_R}\left(\eta_0 - \eta \right)\; \nu - \int_W p e_R \; \nu\right) 
+g \mathcal{V}_Bk \left(\rho_f -\rho_b  \right), \\
\end{align*}
Using~(\ref{eq:mass}) and re-arranging terms, one gets 
\begin{align*}
& \frac{d}{dt} \left(\frac{\rho_f}{2}  \int_{\Sigma_b} r \times \left(n_b \times \nabla \Phi \right) \; \nu+L_b + \rho_f  \int_{\Sigma_f} \Phi n_f  \; \nu \right) \\
&=\lim_{R \rightarrow \infty}  \left( -  \int_{\Sigma_R} p_{atm}n_f  \; \nu + \int_{C_R}\left( p_{atm} -\rho_f  \frac{\nabla \Phi \cdot \nabla \Phi}{2} \right) \; n_s \nu - \int_W p e_R \; \nu  \right)
+g \mathcal{V}_Bk \left(\rho_f -\rho_b  \right), \\
\end{align*}
Using~(\ref{eq:pfarzt}) obtain for the last integral on the right,
\begin{align*}
\lim_{R \rightarrow \infty} \int_W p e_R \; \nu &= \lim_{R \rightarrow \infty} \int_W \left(p_{atm} + \rho_f g (\eta_0-z) + A(R,\theta,z,t) \right) e_R \; \nu. \\
\end{align*} 
The integral of the $A$ term is $O(1/R)$. 
To resolve all the terms containing $p_{atm}$, apply Stokes theorem in $D_R$ to obtain the result 
\begin{align*}
\lim_{R \rightarrow \infty} \int_{D_R} \nabla (1) \; \mu&=0=\lim_{R \rightarrow \infty} \left(\int_{\Sigma_R}  n_f \; \nu- \int_{C_R}  n_s \; \nu  - \int_{\Sigma_b}  n_b \; \nu_b + \int_{W}  e_R \; \nu \right), 
\end{align*}
 where the third integral on the right is zero due to  the closedness of the body. 

Incorporating this result into the linear momentum equation, and since $\int_{W} \rho_f g (\eta_0-z) e_R \; \nu=0$ due to the cylindrical geometry, obtain in the limit $R \rightarrow \infty$, 
\begin{align*}
& \frac{d}{dt} \left(\frac{\rho_f}{2}  \int_{\Sigma_b} r \times \left(n_b \times \nabla \Phi \right) \; \nu+L_b + \rho_f  \int_{\Sigma_f} \Phi n_f  \; \nu \right) \\
&= \int_{\mathcal{S}}\left(-\rho_f  \frac{\nabla \Phi \cdot \nabla \Phi}{2} \right) \; n_s \nu 
+g \mathcal{V}_Bk \left(\rho_f -\rho_b  \right), \\
\end{align*}

Defining 
\begin{align}
\mathcal{L}&=\frac{\rho_f}{2}  \int_{\Sigma_b} r \times \left(n_b \times \nabla \Phi \right) \; \nu+L_b + \rho_f  \int_{\Sigma_f} \Phi n_f  \; \nu 
\end{align}
the linear momentum equation becomes 
\begin{align*}
\frac{d \mathcal{L}}{dt}&=- \rho_f \int_{\mathcal{S}} \frac{\nabla \Phi \cdot \nabla \Phi}{2} n_s \; \nu_s + \left(\rho_f  -\rho_b \right) g \mathcal{V}_Bk.
\end{align*}

\paragraph{Angular Momentum.} Similarly, apply Newton's second law for the evolution of $A_T$,
\begin{align*}
\frac{dA_T}{dt}
&=\lim_{R \rightarrow \infty} \left(- \int_{\Sigma_R} r \times  p_{atm}n_f \; \nu +\int_{C_R} r \times p n_s \; \nu  \right.
\\ & \hspace{1in} \left. - \rho_f g \int_0^\eta \int_{C_R} r \times k  dz \; \nu  + \int_W r \times p e_R \; \nu  \right) \\
& \hspace{2in} +r_c \times \rho_fg \mathcal{V}_Bk - r_c \times \rho_b g \mathcal{V}_Bk ,
\end{align*}
where $r_c$ is the position vector of the centroid of the rigid body. 

 Referring to~(\ref{eq:fluidangmom}) and~(\ref{eq:totmom}), obtain
\begin{align*}
& \frac{d}{dt} \left(- \frac{\rho_f}{2} \int_{\Sigma_b} r^2 \left(n_b \times \nabla \Phi \right) \; \nu + A_b \right) \\
&=\lim_{R \rightarrow \infty} \left(-\rho_f  \frac{d}{dt} \int_{\Sigma_R} r  \times  \Phi n_f \; \nu - \int_{\Sigma_R} r \times  p_{atm}n_f \; \nu + \rho_f  \frac{d}{dt} \int_{C_R} r  \times  \Phi n_s \; \nu \right. \\
& \left. +\int_{C_R} r \times p n_s \; \nu_s   - \rho_f g \int_0^\eta \int_{C_R} r \times k  dz \; \nu + \int_W r \times p e_R \; \nu\right) +r_c \times \left( \rho_f - \rho_b \right) g \mathcal{V}_Bk, \\
&=\lim_{R \rightarrow \infty} \left(-\rho_f  \frac{d}{dt} \int_{\Sigma_R} r  \times  \Phi n_f \; \nu - \int_{\Sigma_R} r \times  p_{atm}n_f \; \nu \right. \\
& \left. +\int_{C_R} r \times \left(\rho_f \frac{\partial \Phi}{\partial t} + p \right)n_s \; \nu   - \rho_f g \int_0^\eta \int_{C_R} r \times k  dz \; \nu + \int_W r \times p e_R \; \nu  \right) +r_c \times \left( \rho_f - \rho_b \right) g \mathcal{V}_Bk, \\
& [{\rm at \; any \;} R \;  {\rm fixed \; in \; time}], \\
&=\lim_{R \rightarrow \infty} \left(-\rho_f  \frac{d}{dt} \int_{\Sigma_R} r  \times  \Phi n_f \; \nu - \int_{\Sigma_R} r \times  p_{atm}n_f \; \nu \right. \\
& \left. +\int_{C_R} r \times \left(p_{atm} + \rho_f g \eta_0 -\rho_f  \frac{\nabla \Phi \cdot \nabla \Phi}{2}  \right)n_s \; \nu   - \rho_f g \int_0^\eta \int_{C_R} r \times k  dz \; \nu + \int_W r \times p e_R \; \nu  \right) \\
& \hspace{3in}+r_c \times \left( \rho_f - \rho_b \right) g \mathcal{V}_Bk, \\
& [{\rm using \;}~(\ref{eq:ps})] \\ 
&=\lim_{R \rightarrow \infty} \left(-\rho_f  \frac{d}{dt} \int_{\Sigma_R} r  \times  \Phi n_f \; \nu - \int_{\Sigma_R} r \times  p_{atm}n_f \; \nu \right. \\
& \left. +\int_{C_R} r \times \left(p_{atm} + \rho_f g (\eta_0 - \eta ) -\rho_f  \frac{\nabla \Phi \cdot \nabla \Phi}{2}  \right)n_s \; \nu_s  + \rho_f g \int_{C_R} r \times \eta n_s \; \nu  - \rho_f g \int_0^\eta \int_{C_R} r \times k  dz \; \nu \right.  \\ 
& \left. \hspace{2in}+ \int_W r \times p e_R \; \nu  \right)  +r_c \times \left( \rho_f - \rho_b \right) g \mathcal{V}_Bk, \\
&=\lim_{R \rightarrow \infty} \left(-\rho_f  \frac{d}{dt} \int_{\Sigma_R} r  \times  \Phi n_f \; \nu - \int_{\Sigma_R} r \times  p_{atm}n_f \; \nu \right. \\
& \left. +\int_{C_R} r \times \left(p_{atm} + \rho_f g (\eta_0 - \eta ) -\rho_f  \frac{\nabla \Phi \cdot \nabla \Phi}{2}  \right)n_s \; \nu + \int_W r \times p e_R \; \nu\right) +r_c \times \left( \rho_f - \rho_b \right) g \mathcal{V}_Bk, \\
&[{\rm since \;} r \times k \; {\rm is \; independent \; of}\; z, \;  \int_0^\eta \int_{C_R} r \times k  dz \; \nu=\int_{C_R} r \times \eta n_s \; \nu] 
\end{align*}
Using~(\ref{eq:pfarzt}) again, 
\begin{align*}
\lim_{R \rightarrow \infty} \int_W r \times p e_R \; \nu &= \lim_{R \rightarrow \infty} \left(\int_{W} r \times \left(p_{atm} + \rho_f g (\eta_0-z) + A(R,\theta,z,t) \right) e_R \; \nu \right),
\end{align*}
The integral of the $A$ term is again $O(1/R)$, since $r \times e_R$ filters off the $R$-coordinate. 

Now use another result obtained from Stokes' theorem applied in $D_R$
\begin{align*}
\int_{D_R} \nabla \times  r \; \mu &=0=  \int_{\Sigma_R} n_f \times  r \; \nu - \int_{C_R} n_s \times  r \; \nu -\int_{\Sigma_b} n_b \times  r \; \nu \\
& \hspace{3in} + \int_{W} e_R \times r\; \nu_R
\end{align*}
The third integral can be shown to be equal to zero by applying the same integral theorem again with the body as the domain.  Incorporating this result into the angular momentum equation, and noting that due to the cylindrical geometry of $W$, 
\[\int_{W} r \times \rho_f g (\eta_0-z) e_R \; \nu = \int_W \rho_f g z (\eta_0-z) e_t \; \nu=0, \]
\begin{align*}
& \frac{d}{dt} \left(- \frac{\rho_f}{2} \int_{\Sigma_b} r^2 \left(n_b \times \nabla \Phi \right) \; \nu + A_b + \rho_f \int_{\Sigma_f} r  \times  \Phi n_f \; \nu \right) \\
&=\lim_{R \rightarrow \infty} \left(\int_{C_R} r \times \left(\rho_f g (\eta_0 - \eta ) n_s - \rho_f  \frac{\nabla \Phi \cdot \nabla \Phi}{2}  \right)\; \nu_s \right)  +r_c \times \left( \rho_f - \rho_b \right) g \mathcal{V}_Bk
\end{align*}

 Invoking~(\ref{eq:angmass}) and defining 
\begin{align*}
\mathcal{A}&= - \frac{\rho_f}{2} \int_{\Sigma_b} r^2 \left(n_b \times \nabla \Phi \right) \; \nu + A_b + \rho_f \int_{\Sigma_f} r  \times  \Phi n_f \; \nu
\end{align*}
obtain in the limit $R \rightarrow \infty$, 
\begin{align*}
\frac{d \mathcal{A}}{dt}&=\int_{\mathcal{S}} r \times \left(- \rho_f  \frac{\nabla \Phi \cdot \nabla \Phi}{2}  \right)\; \nu_s +r_c \times \left( \rho_f - \rho_b \right) g \mathcal{V}_Bk
\end{align*}

\newpage 

\section{Appendix B: Global momentum evolution equations in a body-fixed frame.}

\paragraph{Linear momentum.} Using again the relations~(\ref{eq:rlrel})--~(\ref{eq:gradftrans}) and the change of variables theorem, the linear momentum $\mathcal{L}$ transforms as 
\begin{align*}
\mathcal{L}&=\frac{\rho_f}{2}  R(t) \int_{\bar{\Sigma}_b} l \times \left(\bar{n}_b \times \nabla_b \Phi' \right) \; \bar{\nu}+R(t) \left(\bar{b}(t) \times \int_{\bar{\Sigma}_b} \left(\bar{n}_b \times \nabla_b \Phi_b' \right) \; \bar{\nu} \right) \\
& \hspace{2in} + R(t) \bar{L}_b + \rho_f  R(t) \int_{\bar{\Sigma}_f} \Phi' \bar{n}_f  \; \bar{\nu}. 
\end{align*}
Next, show that 
\[ \int_{\bar{\Sigma}_b} \left(\bar{n}_b \times \nabla_b \Phi' \right) \; \bar{\nu}=0\]
by using the following argument. Let $\bar{\Phi}_b'$ satisfy the following Dirichlet problem
\begin{align*}
\nabla^2 \bar{\Phi}_b'&=0 \quad {\rm in } \; B, \\
\bar{\Phi}_b'&=\Phi_b' \quad {\rm in} \; \Sigma_b.
\end{align*}
Then 
\begin{align}
\int_{\bar{\Sigma}_b} \left(\bar{n}_b \times \nabla_b \Phi_b' \right) \; \bar{\nu}&=\int_{\bar{\Sigma}_b} \left(\bar{n}_b \times \nabla_b \bar{\Phi_b}' \right) \; \bar{\nu}=\int_B \nabla_b \times \nabla_b \bar{\Phi_b}' \; \bar{\mu}=0, \label{eq:result}
\end{align}
and therefore 
\begin{align*}
\mathcal{L}&=  R(t) \left(\frac{\rho_f}{2} \int_{\bar{\Sigma}_b} l \times \left(\bar{n}_b \times \nabla_b \Phi' \right) \; \bar{\nu} + \bar{L}_b + \rho_f  \int_{\bar{\Sigma}_f} \Phi' \bar{n}_f  \; \bar{\nu} \right), \\
&=: R(t) \mathfrak{L}  
\end{align*}
where $\bar{L}_b=\rho_b m_b \bar{V}$. The linear momentum equation in $\mathfrak{L}$ becomes 
\begin{align*}
\frac{d \mathfrak{L}}{dt}+ \bar{\Omega} \times \mathfrak{L}&=0
\end{align*}

\paragraph{Angular momentum.} Similarly, transform the angular momentum $\mathcal{A}$, 
\begin{align*}
\mathcal{A}&= - \frac{\rho_f}{2} \int_{\Sigma_b} r^2 \left(n_b \times \nabla \Phi \right) \; \nu + A_b + \rho_f \int_{\Sigma_f} r  \times  \Phi n_f \; \nu, \\
&=- \frac{\rho_f}{2} \int_{\Sigma_b} r^2 \left(n_b \times \nabla \Phi \right) \; \nu +b \times M_b V + I \Omega  + \rho_f \int_{\Sigma_f} r  \times  \Phi n_f \; \nu,  \\
&[{\rm using ; \ref{eq:bodymomsp}}] 
\end{align*}
Proceeding, using relations~(\ref{eq:rlrel})--~(\ref{eq:gradftrans}) and the change of variables theorem,
\begin{align*}
\mathcal{A}&= - \frac{\rho_f}{2} R(t) \int_{\bar{\Sigma}_b} \left(l^2 + 2 l \cdot \bar{r}_c + \bar{r}_c^2 \right) \left(\bar{n}_b \times \nabla_b \Phi' \right) \; \bar{\nu} +R(t) \left(\bar{r}_c \times M_b \bar{V} + \bar{I {\Omega}} \right) \\
& \hspace{2in} + \rho_f R(t) \int_{\bar{\Sigma}_f} \left(l  + \bar{r}_c \right) \times  \Phi' \bar{n}_f \; \bar{\nu},  \\
&= - \frac{\rho_f}{2} R(t) \int_{\bar{\Sigma}_b} \left(l^2 + 2 l \cdot \bar{r}_c \right) \left(\bar{n}_b \times \nabla_b \Phi' \right) \; \bar{\nu} +R(t) \left(\bar{r}_c \times M_b \bar{V} +\bar{ I {\Omega}} \right) \\
& \hspace{2in} + \rho_f R(t) \int_{\bar{\Sigma}_f} \left(l  + \bar{r}_c \right) \times  \Phi' \bar{n}_f \; \bar{\nu}, \\
&[{\rm using \; result~(\ref{eq:result})}] \\
&= R(t) \left( - \frac{\rho_f}{2} \int_{\bar{\Sigma}_b} l^2 \left(\bar{n}_b \times \nabla_b \Phi' \right) \; \bar{\nu} + \bar{I \Omega} + \rho_f \int_{\bar{\Sigma}_f} l  \times  \Phi' \bar{n}_f \; \bar{\nu} \right)\\
& \hspace{1in} - \frac{\rho_f}{2} R(t) \int_{\bar{\Sigma}_b} 2 l \cdot \bar{r}_c \left(\bar{n}_b \times \nabla_b \Phi' \right) \; \bar{\nu} + R(t) \left(\bar{r}_c \times M_b \bar{V}  \right)+  \rho_f R(t) \int_{\bar{\Sigma}_f} \bar{r}_c \times  \Phi' \bar{n}_f \; \bar{\nu}. 
\end{align*}
At this point, a Proposition proved in \cite{ShShKeMa2008} is invoked (adapted to the notation of this paper): 
\begin{prop} For any gradient vector field $\nabla f$ defined on $\Sigma_b$, the following is true:
\begin{align*}
\int_{\Sigma_b}  (b \cdot  l ) n_b \times \nabla f \; \nu &= - \int_{\Sigma_b} b \cdot \left( n_b \times \nabla f \right) l  \; \nu. 
\end{align*}
\end{prop}
where $b$ is a constant vector. This holds in either a spatially-fixed or a body-fixed frame. Therefore, 
\begin{align*}
\mathcal{A}&= R(t) \left( - \frac{\rho_f}{2} \int_{\bar{\Sigma}_b} l^2 \left(\bar{n}_b \times \nabla_b \Phi' \right) \; \bar{\nu} +  \bar{I \Omega} + \rho_f  \int_{\bar{\Sigma}_f} l  \times  \Phi' \bar{n}_f \; \bar{\nu} \right)\\
& \hspace{1in} + R(t)  \left[\bar{r}_c \times \frac{\rho_f}{2} \int_{\bar{\Sigma}_b}  l \times \left(\bar{n}_b \times \nabla_b \Phi' \right) \; \bar{\nu} + \bar{r}_c \times M_b \bar{V}  +   \bar{r}_c \times  \rho_f \int_{\bar{\Sigma}_f} \Phi' \bar{n}_f \; \bar{nu}, \right]\\
&= R(t) \left( - \frac{\rho_f}{2} \int_{\bar{\Sigma}_b} l^2 \left(\bar{n}_b \times \nabla_b \Phi' \right) \; \bar{\nu} + \bar{I \Omega} + \rho_f  \int_{\bar{\Sigma}_f} l  \times  \Phi' \bar{n}_f \; \bar{\nu} \right) + R(t)  \left[\bar{r}_c \times \mathfrak{L} , \right]
\end{align*}
Defining
\begin{align*}
\mathfrak{A}&=- \frac{\rho_f}{2} \int_{\bar{\Sigma}_b} l^2 \left(\bar{n}_b \times \nabla_b \Phi' \right) \; \bar{\nu} + \bar{I \Omega} + \rho_f  \int_{\bar{\Sigma}_f} l  \times  \Phi' \bar{n}_f \; \bar{\nu}, 
\end{align*}
the angular momentum equation transforms as 
\begin{align*}
R(t) \left( \frac{d \mathfrak{A}}{dt} + \bar{\Omega} \times \mathfrak{A} +  \bar{V} \times \mathfrak{L} \right)&=0, \\
\Rightarrow \frac{d \mathfrak{A}}{dt} + \bar{\Omega} \times \mathfrak{A} +  \bar{V} \times \mathfrak{L}&=0
\end{align*}
Note that in the above, 
\[R(t)\bar{V}=V \equiv dr_c/dt, \quad \bar{V} \neq d \bar{r}_c/dt\]

\end{document}